# Structurally induced magnetic transitions in layered dichalcogenides $MoQ_2$ (Q = S, Se, Te) and double hydroxides $(M^{2+})_6Al_3(OH)_{18}[Na(H_2O)_6](SO_4)_2$ $6H_2O$ ($M^{2+}$ = $Mn^{2+}$, $Fe^{2+}$) under mechanical deformation

**L M Volkova**

Institute of Chemistry, Far Eastern Branch, Russian Academy of Sciences, 690022 Vladivostok, Russia

E-mail: volkova@ich.dvo.ru

Exploring how mechanical strain can modify the magnetic properties of low-dimensional structures is one of the priorities of straintronics, an area in condensed matter physics. It has been proven by calculating the parameters of magnetic interactions $J_{ij}$ and developing structural/magnetic models of the layered dichalcogenides $MoS_2$, $MoSe_2$, $MoTe_2$ and layered double hydroxides $(M^{2+})_6Al_3(OH)_{18}[Na(H_2O)_6](SO_4)_2$ $6H2O$ ($M^{2+}$ = $Mn^{2+}$, $Fe^{2+}$) with a grapheme type structure that magnetic interactions are responsive to the mechanical deformation of their crystal structure. As turned out, the ions in these antiferromagnetic materials are situated in the hexagonal planes close to critical positions. We have thus demonstrated that the fluctuations of the intermediate ions near critical positions due to mechanical strain cause dramatic changes to the magnetic parameters and allow the magnetic properties to be modified by mechanical strain. To be sure, an abundant class of new 2D materials transition-metal-based double hydroxides, whose properties are similar to those of molybdenum-based chalcogenides have promise as materials to be used in straintronics.

______________________________________________________________________

## 1. Introduction

Exploring how mechanical strain can modify the magnetic properties of low-dimensional structures is one of the priorities of straintronics, a new research area in condensed matter physics [1, 2]. Magnetic straintronics becomes a platform for new-generation data-processing hardware to be developed. The relationship between mechanical stresses and the magnetic subsystem, as well as changes in the crystal structure under mechanical deformations, have been confirmed using strain engineering methods and physical effects caused by deformation in solids [1–5]. This allows us to consider the crystal structure as an intermediary linking mechanical deformation and the magnetic subsystem.

Mechanical deformation in magnetic films or particles can be attained in a variety of ways [2], for example, directly, by bending or by using a substrate that have other lattice parameters. Another simplest way is by bending the substrate on which the material of interest is deposited or by depositing films or particles on previously bent substrates. According to [4, 5, 6], deformation in two-dimensional van der Waals materials leads to changes not only in the strength of magnetic moments, but also in spin ordering.

We will be treating mechanical deformation as a phenomenon that relates to atomic motions – just in the way it is treated when exploring plasticity [7, 8]. Magnetic phase transitions serve as evidence that magnetic interactions can be responsive to the mechanical deformation of the crystal structure, this effect being known as the 'magnetic deformation effect' (MDE).

The role of mechanical deformation is in initiating spin transitions within a pair of interacting magnetic ions followed by having them assume a more favorable spin configuration. Deformation by mechanical processes can switch spin transitions from the paramagnetic to the magnetically ordered state – or from ferromagnetic to antiferromagnetic. Thin two-dimensional materials are flexible, easy to expand or compress, with interatomic distances and angles changing their values.

The main objective of our study was to find out as to whether there was an association between the mechanical deformation of the crystal structure of layered magnetic compounds on the one hand, and magnetic transitions and anomalous changes in the strength of magnetic interactions on the other. Because the strength of magnetic interactions and the magnetic ordering types largely depend on the geometrical arrangement and size of intermediate ions in the local spaces between magnetic ions and the distances between magnetic ions, the goal was considered to be attainable [9, 10]. In particular, the dependence of the nearest-neighbor exchange interactions on the M-X-M bond angle is widely used for prediction of the sign and relative strength of magnetic interactions (Goodenough–Kanamori–Anderson's empirical rules [11-13]).

To this end, we calculated the parameters (sign and strength) of the magnetic interactions $J_{ij}$ in the layered dichalcogenides $MoS_2$, $MoSe_2$, $MoTe_2$ and the layered double hydroxides $(M^{2+})_6Al_3(OH)_{18}[Na(H_2O)_6](SO_4)_2\ 6H_2O$ ($M^{2+} = Mn^{2+}$, $Fe^{2+}$) possessing a graphene-type crystal structure or fragments of suchlike and developed their structural/magnetic models. There were two factors supporting the rationale behind the choice of compounds. First, the magnetic interactions are strong in the planes of magnetic ions and infinitesimally weak between the planes due to large distances between the planes. Secondly, even a minor displacement of intermediate ions in the local spaces of magnetic interactions may lead to a sharp change in the strength of the magnetic interactions or to AFM↔FM transitions in the planes of the magnetic ions. Smooth AFM-FM transitions are impossible, because the mechanical deformation of the crystal makes the intermediate ions in the unstable domain leave the critical positions in jerks. The pressure-driven metal-nonmetal transition occurs extremely rapidly once the critical value $a_0$ in the unstable domain is reached [14].

The results of this study are most relevant to a structural data-based search for new 2D materials with magnetic ordering easily modifiable via altering their crystal structure by mechanical deformation.

## 2. Method of calculation

Structural/magnetic models are based on crystal chemical parameters (crystal structure, ion charge and ion size). The variables of these models include: (1) the sign and strength of the magnetic interactions $J_{ij}$; (2) the dimensionality of magnetic structures (this is not always the same as the dimensionality of the crystal structures); (3) the presence of magnetic frustrations in specific geometric configurations; (4) a possibility to reorient magnetic moments (that is, to enable AFM-to-FM transitions) due to the departure of the intermediate ions from critical positions.

To infer the sign (type) and strength of the magnetic interactions $J_{ij}$ from structural data, we used the Crystal Chemistry Method, our previous development, and the associated software program MagInter [9-10]. The Crystal Chemistry Method puts together three well-known concepts about the nature of magnetic interactions: Kramers's idea [15], the Goodenough–Kanamori–Anderson's model [11-13] and the polar Shubin–Vonsovsky's model [16].

The Crystal Chemistry Method allows the sign (type) and strength of the magnetic interactions $J_{ij}$ to be inferred from structural data. According to this method, a coupling between the magnetic ions $M_i$ and $M_j$ occurs at the moment an intermediate ion $A_n$ crosses the boundary between them with an overlap of ~0.1 Å (Figure 1(a) and (b)). The area of the local space between the $M_i$ and $M_j$ ions along the bond line is defined as a cylinder, whose radius is equal to the radius of any of these ions. The strength of the magnetic couplings and the type of magnetic ordering in insulators are determined mainly by the geometrical position and the size of intermediate ions $A_n$ in the local space between two magnetic ions ($M_i$ and $M_j$).

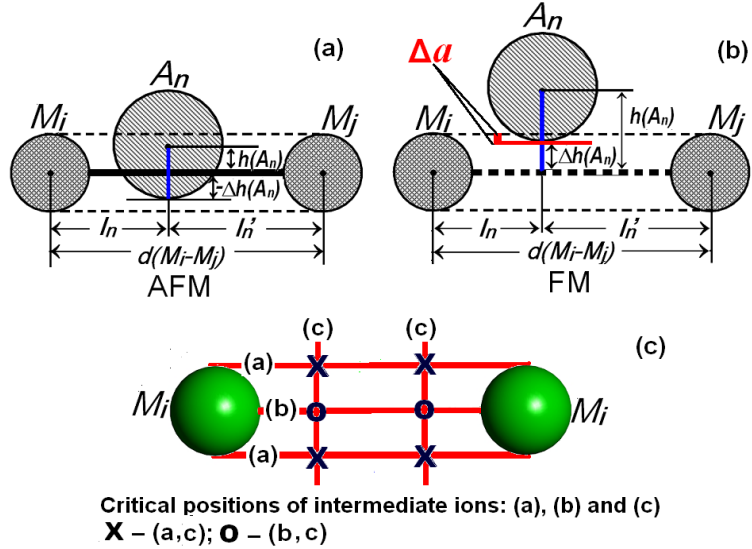

**Fig. 1.** Arrangement of the intermediate ions *An* in the local space between the magnetic ions *Mi* and *Mj* in the cases when *An* initiates (**a**) antiferromagnetic and (**b**) ferromagnetic interactions. *h(An)*, *ln*, *ln'*, and *d(Mi–Mj)* are the parameters that account for the sign and strength of the magnetic interactions *Jn*. (**c**) Critical positions of the intermediate ions: (a) $h(A_n) = r_M + r_{An}$, (b) $h(A_n) = rA_n$ ($h(A_n) = 0$), (c) $l_n'/l_n = 2$.

The positions of the intermediate ion $A_n$ in the local space are determined (1) by the distance $h(A_n)$ between the center of the ion $A_n$ and the $M_i$-$M_j$ bond line and (2) by the shift of the intermediate ion towards one of the magnetic ions expressed as the ratio of length $l_n$ to length $l_n'$. This ratio explains the departure of the intermediate ion along the $M_i$-$M_j$ bond line from the center point between the magnetic ions ($l_n \leq l_n'$; $l_n' = d(M_i - M_j) - l_n$) (Figure 1 (a) and (b)).

The intermediate ions $A_n$ will tend to orient the magnetic moments of the $M_i$ and $M_j$ ions and make their contributions $j_n$ to the emergence of the AFM or FM components of the magnetic interaction depending on (1) the overlap, $\Delta h(A_n)$, of the local space between magnetic ions, (2) the asymmetry ($l_n'/l_n$) of the position relative to the central $M_i$-$M_j$ bond line, and (3) the distance, $M_i$-$M_j$, between the magnetic ions. From among the above parameters, only the overlap between the magnetic ions $M_i$ and $M_j$ ($\Delta h(A_n) = h(A_n) - r_{A_n}$) is equal to the difference between the distance $h(A_n)$ from the center of the ion $A_n$ to the $M_i$-$M_j$ bond line and the radius ($r_{A_n}$) of the ion $A_n$ determines the sign of the magnetic interaction. If $\Delta h(A_n) < 0$, there is a $|\Delta h|$-wide overlap between the ion $A_n$ and the $M_i$-$M_j$ bond line, and so the intermediate ion starts contributing to the AFM component of the magnetic interaction. If $\Delta h(A_n) > 0$, there is a $\Delta h$-wide gap between the bond line and the ion $A_n$, and this ion starts contributing to the FM component of the magnetic interaction.

The contribution $j_n$ is defined as

$$j_n = \frac{\Delta h(A_n)\frac{l_n}{l_n'} + \Delta h(A_n)\frac{l_n'}{l_n}}{d(M_i - M_j)^2} \quad \text{(if } l_n'/l_n < 2.0\text{)}, \quad (1)$$

and

$$j_n = \frac{\Delta h(A_n)\frac{l_n}{l_n'}}{d(M_i - M_j)^2} \quad \text{(if } l_n'/l_n \geq 2.0\text{)}. \quad (2)$$

The sign and strength of the magnetic coupling $J_{ij}$ are determined by the sum of the above contributions:

$$J_{ij} = \sum_n j_n \quad (3)$$

$J_{ij}$ is expressed in per angstrom units (Å$^{-1}$). If $J_{ij} < 0$, the magnetic ordering of the ions $M_i$ and $M_j$ is antiferromagnetic, while if $J_{ij} > 0$, it is ferromagnetic, and if $J_{ij} = 0$ the system undergoes a transition to the paramagnetic state.

By looking at Eqs. (1) - (3) it is possible to see why aberrant magnetic interactions and magnetic phase transitions take place in magnets. There are several critical positions of the intermediate ions $A_n$ such that even a slight departure from them could lead to the reorientation of magnetic moments (that is, to an AFM–FM transition) and/or a dramatic change in the strength of the magnetic interaction.

Noteworthy, ions in a crystal structure can be displaced due to such factors as temperature, mechanical deformation, pressure and magnetic field, to mention a few. That is why when predicting possible changes in the sign and strength of magnetic interactions, not only the ions at critical positions should be taken into account, but also those in their vicinity (Figure 1(c)). In the compounds considered in this work, the intermediate ions take on the following critical positions:

(a) $h(A_n) = r_M + r_{An}$ : the distance $h(A_n)$ from the center of the ion $A_n$ to the $M_i$–$M_j$ bond line is equal to the sum of the respective radii of the ions $M$ and $A_n$. The ion $A_n$ reaches the surface of a cylinder with radius $r_M$, limiting the space area between the magnetic ions $M_i$ and $M_j$. In this case, the ion $A_n$ does not induce magnetic interaction. However, the slightest decrease in $h(A_n)$ (that is, a displacement of the ion $A_n$ within this area) leads to a strong FM interaction between the magnetic ions.

(b) $h(A_n) = rA_n$ ($h(A_n) = 0$): the distance $h(A_n)$ from the center of the ion $A_n$ to the $M_i$–$M_j$ bond line is equal to the radius of the ion $A_n$ ($A_n$ reaches the $M_i$–$M_j$ bond line). In this case, the magnetic fields stop interacting. However, a slight decrease in $h(A_n)$, meaning that the ion $A_n$ overlaps the $M_i$–$M_j$ bond line, leads to a weak AFM interaction, while a slight increase in $h(A_n)$, meaning that there is a gap between the ion $A_n$ and the $M_i$–$M_j$ bond line, leads to a weak FM interaction.

(c) $l_n'/l_n = 2$: the insignificant displacement (up to $l_n'/l_n < 2.0$) of an ion $A_n$ to the centre between magnetic ions in parallel by the line connecting $M_i$-$M_j$ results in a dramatic increase in the strength of magnetic interaction.

Next, we will show that the structural/magnetic models built on the basis of magnetic coupling parameters computed by the Crystal Chemistry Method make it possible to reveal the main correlations between the crystal structure of compounds and their magnetic properties. Displacement of intermediate ions from critical positions by mechanical deformation is normally accompanied by magnetic transitions, FM to AFM, FM to PM or AFM to PM.

The input data format for the MagInter software program (crystallographic parameters, atom coordinates) is compatible with CIF files in the Inorganic Crystal Structure Database (ICSD) (FIZ Karlsruhe, Germany). The ionic radii of Shannon [17]: r($^{VI}$Mo$^{4+}$) = 0.65 Å, r($^{VI}$S$^{2-}$) = 1.84 Å, r($^{VI}$Se$^{2-}$) = 1.98 Å, r($^{VI}$Te$^{2-}$) = 2.21 Å, r($^{VI}$Fe$^{2+}$) = 0.78 Å, r($^{VI}$Mn$^{2+}$) = 0.83 Å, r($^{VI}$O$^{2-}$) =

1.40 Å, r($^{IV}S^{6+}$) = 0.12 Å were used for calculations. Tables 1, 2, 3 and 4 (Supplementary Note 1) show the crystallographic characteristics and parameters of the magnetic interactions $J$n calculated on the basis of structural data and the distances between the magnetic ions in the materials under study. Additionally, the overlap of the local spaces between magnetic ions (Δh(X)), the asymmetry (l′n/ln) of the position relative to the central $M_i$–$M_j$ bond line, and the $M_i$–X–$M_j$ angle are presented for the intermediate ions X, which provide the maximum contributions ($j$(X)) to the AFM or FM components of these couplings $J$n.

There is quite a literature on the synthesis and structural studies of the molybdenum dichalcogenides $MoQ_2$. Little difference was observed in the crystallographic parameters and atom coordinates between the samples of these compounds sharing the same composition and crystallizing in the same space group. With the Crystal Chemistry Method, we calculated the parameters of the magnetic interactions $J_{ij}$ in 18 dichalcogenide and two hydroxide samples. The samples represent:

(1) centrosymmetric samples with space group P63/mmc (N194) 2H-$MoS_2$ (ICSD: 24000, 105091, 644245, 644246, 644250, 644259), 2H-$MoSe_2$ (ICSD: 191306, 049800, 644335) and 2H-$MoTe_2$ (ICSD: 15431, 24155, 644476, 64481);

(2) non-centrosymmetric samples with space group R3m (N160) 3H-$MoS_2$ (ICSD: 38401, 76370, 644249) and 3R-$MoSe_2$ (ICSD 016948);

(3) a sample with the centrosymmetric monoclinic space group P2$_1$/m (N11) $MoTe_2$ (ICSD 14349);

(4) two layered double hydroxides with the centrosymmetric trigonal/rhombohedral space group R-3 (N148): shigaite [AlMn$^{2+}_2$(OH)$_6$]$_3$(SO$_4$)$_2$Na(H$_2$O)$_6$\{H$_2$O\}$_6$ (ICSD 82492) and nikischerite [AlFe$^{2+}_2$(OH)$_6$]$_3$(SO$_4$)$_2$Na(H$_2$O)$_6$\{H$_2$O\}$_6$ (ICSD 97312).

Tables 1, 2, 3 and 4 (Supplementary Note 1) show the crystallographic characteristics and parameters of the magnetic couplings $J$n calculated on the basis of structural data and the distances between the magnetic ions in the materials under study. Additionally, the overlap of the local spaces between magnetic ions (Δh(X)), the asymmetry (l′n/ln) of the position relative to the central Mi–Mj bond line, and the Mi–X–Mj angle are presented for the intermediate ions X, which provide the maximum contributions ($j$(X)) to the AFM or FM components of these couplings $J$n.

# 3. Results and discussion

3.1. How to regulate the magnetic properties of the low-dimensional structures of molybdenum-based dichalcogenides and iron- and manganese-based double hydroxides

Exchange magnetic interactions are more sensitive to geometrical changes in low-dimensional than high-dimensional magnetic systems. We have developed structural/magnetic models of the two-dimensional dichalcogenides $MoQ_2$ (Q = S, Se, Te) and double hydroxides $(M^{2+})_6Al_3(OH)_{18}[Na(H_2O)_6](SO_4)_2 \cdot 6H_2O$ ($M^{2+}$ = $Mn^{2+}$, $Fe^{2+}$). With these models, we will now demonstrate that transitions of magnetic interactions from a ferromagnetic to an antiferromagnetic state are a possibility, and so are the changes in the strength of these magnetic interactions even after minor displacements of ions caused by mechanical tension. Exchange magnetic interactions are sensitive to out-of-plane displacements of the chalcogenide atoms (S, Se and Te), which act as intermediate ions in exchange interactions, and to an in-plane displacements of the Mo atoms, that is, the magnetic ions $Mo^{4+}$. Furthermore, we will show what effect an increase in the size of the intermediate ion Q, an increase in the number of $MoQ_2$ layers as well as contraction and expansion of the unit cell of layered compounds have on the parameters of magnetic interactions.

3.1.1. The crystal structure of the molybdenum dichalcogenides $MoQ_2$ (Q = S, Se, Te)

The molybdenum dichalcogenides $MoS_2$ [18-21], $MoSe_2$ [22, 23, 24] and $MoTe_2$ [23, 25] have once become known for being strongly anisotropic. These compounds appear as layers with strong covalent bonds within them, while the neighboring layers are bonded by weak van der Waals interactions. A molybdenum dichalcogenide layer is a monolayer of hexagonally packed metal atoms sandwiched between two planes of chalcogenide atoms (Fig. 2). The crystal structure of the molybdenum dichalcogenide appears as a result of hexagonal stacking in Q-Mo-Q sequence without displacement. In each Q-Mo-Q sandwich layer, the molybdenum atom is bonded to the six nearest-neighboring Q atoms in trigonal prismatic coordination. In this paper, we will consider the $2H-MoQ_2$ and $3R-MoQ_2$ polytypes. The number two in $2H-MoQ_2$ and the number three in $3R-MoQ_2$ tell us how many $MoQ_2$ molecules the unit cell has in it. It is obvious that the edge $c$ of the unit cell of 3R-MoQ$_2$ is about 1.5 times as great as that of 2H-MoQ$_2$, with $a$ being virtually the same. Natural or synthetic MoS$_2$ may appear in the form of the 2H or 3R polytypes – or a mix of both. In nature, 2H-MoS$_2$ is prevalent.

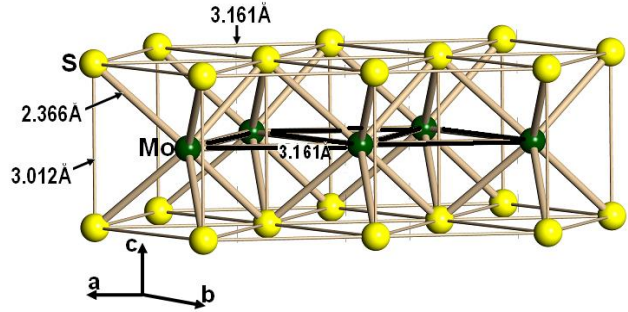

Fig. 2. A layer of the molybdenum dichalcogenide MoS$_2$.

L.C Towle et al. [21] showed that the three-layer rhombohedral form of the molybdenum diselenide 3R-MoSe$_2$ can be produced by exposure of the normal two-layer hexagonal 2H-MoSe$_2$ to a pressure as high as 40 kilobar and a temperature of 1500 °C. The new form is isostructural with the rhombohedral molybdenum disulfide MoS$_2$.

A high-temperature polymorph of MoTe$_2$ [25] crystallizes in the centrosymmetric monoclinic space group P2$_1$/m. The crystal structure of this polymorph appears as Te-Mo-Te sandwiches, as do the other molybdenum dichalcogenides. However, here the Mo atoms of the adjacent MoTe$_6$ octahedra are closer to each other. Thus, each Mo atom has eight neighbors: six Te atoms and two Mo atoms. This off-center position of the metal atoms in the tellurium octahedra changes dramatically the parameters of the magnetic interactions $J$ij in the high-temperature modification of MoTe$_2$.

## 3.1.2. Exchange magnetic interactions are responsive to displacements of intermediate ions

In the two-layer hexagonal $P6_3/mmc$ (N194) 2H-MoQ$_2$ (Q = S, Se, Te) (Fig. 3) and three-layer rhombohedral $R3m$ (N160) 3R-MoQ$_2$ (Q = S, Se) (Fig. 4) modifications, the magnetic ions Mo$^{4+}$ form 2D triangular lattices composed of edge-sharing triangles, each spin having six nearest neighbors. In these triangular lattices, the parameters of the nearest-neighbor magnetic interaction $J$1 and the next-nearest-neighbor magnetic interaction $J$3 change dramatically following a minor displacement of the intermediate ions Q. For details, see Supplementary Note 1, Fig. 5 and Tables 1 and 2.

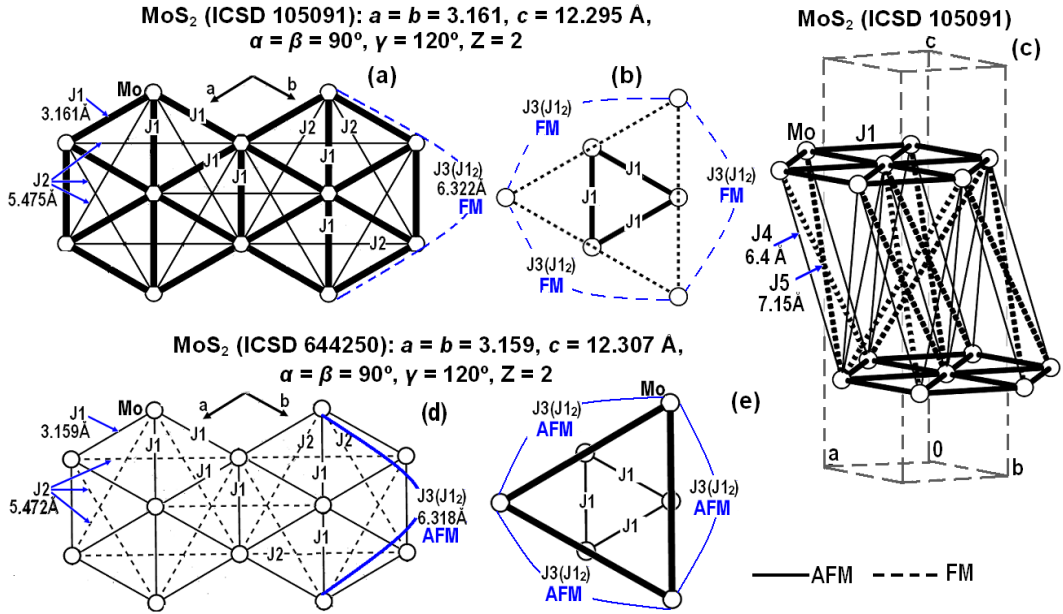

Fig. 3. Interactions Jn in two two-layer hexagonal MoS$_2$ samples: (I) ICSD 105091 (a), (b), (c) and (II) ICSD 644250 (d), (e). In this and other figures, the width of the lines reflects the amount of strength of the interactions Jn. Antiferromagnetic (AFM) and ferromagnetic (FM) interactions are indicated as solid and dashed lines, respectively.

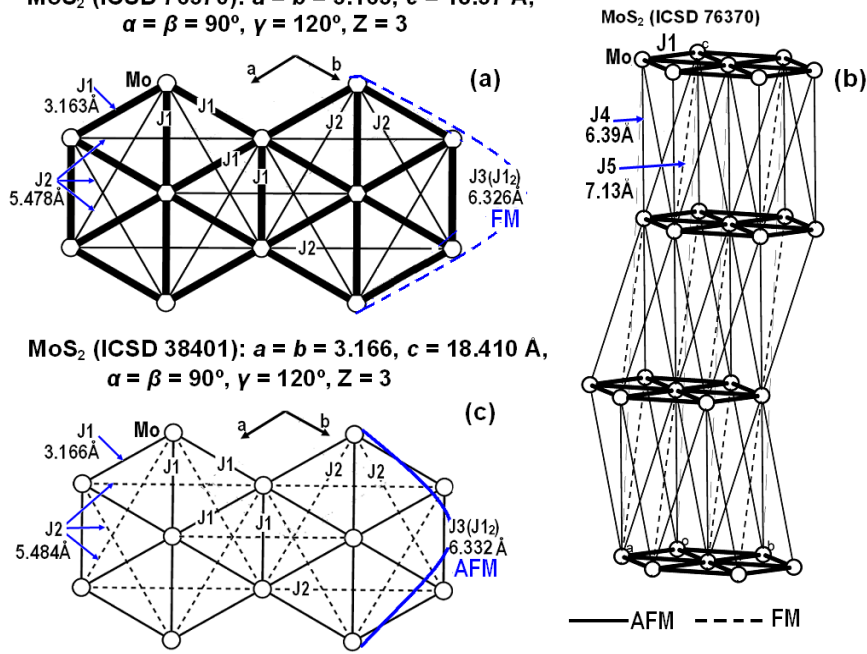

Fig. 4. Interactions Jn in two three-layer samples, 3R-MoS$_2$ (ICSD-76370) and 3R-MoS$_2$ (ICSD-38401).

The exchange system is unstable because the local space of interactions between the tiny magnetic ions Mo is small and the intermediate ions Q are large. As a result, the intermediate ions Q occur at two critical positions at once (Fig. 1(c)): (*a*) near the boundaries of the local space and (b) near the central Mo-Mo bond line, and the least displacement may lead to the reorientation of magnetic moments (that is, to AFM-to-FM transitions) and/or a dramatic change in the strength of the magnetic interaction.

Two 2H-$MoS_2$ samples in the Inorganic Crystal Structure Database perfectly exemplify this: sample I (ICSD 105091) [18] (Supplementary Note 1, Figure 5(a) and (b)) and sample II (ICSD 644250) [19] (Figure 5(c) and (d)), which have disparate strengths of the main magnetic interactions in the *ab* plane. According to our calculations, in sample I, the strong AFM nearest-neighbor couplings $J1$ ($J1$ = -0.0316 Å$^{-1}$, d(Mo1-Mo1) = 3.161 Å) between the Mo1 ions are dominating in the triangular lattice and compete with each other in the smaller triangles. The third longest $J3(J1_2)$ is quite a strong FM interaction ($J3(J1_2)/J1$ = -0.55) and is not competing with the nearest-neighbor AFM interaction $J1$ (↑↓↑) in the chains along the sides of the smaller triangles. Additionally, these FM interactions $J3(J1_2)$ are not competing with each other in the larger triangles FM $J3(J1_2)$ - FM $J3(J1_2)$ - FM $J3(J1_2)$.

However, AFM $J_1$ and FM $J3(J1_2)$ are unstable interactions. The difference between sample II (ICSD 644250) (Figure 5(c) and (d)) and sample I (ICSD 105091) is tenuous: just a 0.07-Å displacement of the intermediate ion S1 from the Mo-Mo bond line in the direction of the boundary of the local space. These tiny structural changes led to an abrupt 8.8-fold decrease in AFM $J1$ in sample II (Figure 5(c)) compared to sample I. However, on the other hand, this displacement leads to an FM $J3(J1_2)$ → AFM $J3(J1_2)$ transition (Figure 5(b) and (d)) and weak AFM $J1$ cannot compete against strong AFM $J3(J1_2)$ ($J1/J3(J1_2)$ = 0.1) any further. Nevertheless, sample II exhibits frustration in the larger AFM triangles (Figure 3(d) and (e)), in which the AFM interactions $J3(J1_2)$ are in competition with each other.

This example shows us that the intermediate ions Q (S, Se and Te) fluctuating around the critical positions lead to sharp changes in magnetic parameters, thus allowing the magnetic properties to be modified by mechanical deformation. Figures (3) and (4) show how minor changes to structural parameters cause major changes in magnetic interactions.

### 3.1.3. Out-of-plane magnetic interactions weaken as MoS$_2$ layers grow in number

We explored how the exchange magnetic interactions depend on the number of MoQ$_2$ layers. As the number of the MoS$_2$ layers changes from 2 to 3, the structure switches from centrosymmetric hexagonal (space group P6$_3$/mmc (N194)) to non-centrosymmetric trigonal (space group R3m (N160)).

However, according to our calculations (Figures 3 and 4, Supplementary Note 1, Tables 1 and 2), this transition has little effect, if any, on the parameters of the in-plane magnetic interactions $J$1 and $J$2. The parameters of these interactions in the pairs composed of the two-layer and three-layer samples, such as 2H-MoS$_2$ (ICSD-105091) - 3R-MoS$_2$ (ICSD-76370) [18] and 2H-MoS$_2$ (ICSD-644250) [19] - 3R-MoS$_2$ (ICSD-38401) [20], are similar. The difference between the out-of-plane AFM interactions $J$4 in these samples is insignificant, too, while the FM interactions $J$5 are substantially, 4- to 5-fold, weaker in the three-layer than two-layer samples. The weakening of the out-of-plane FM interactions $J$5 following the increase in the number of MoS$_2$ layers to 3 makes the main difference between 2H-MoS$_2$ and 3R-MoS$_2$. This change virtually eliminates the competition between the weak out-of-plane FM interactions $J$5 and the strong in-plane AFM interactions $J$3 in the triangles FM$J$5-FM$J$5-AFM$J$3 ($J$5/$J$3 = -0.19) of the 3R-MoS$_2$ sample (ICSD-38401). A similar 4.5-fold decrease in FM interactions $J$5 is observed as 2H-MoSe$_2$ switches to 3R-MoSe$_2$.

### 3.1.4. The size of the intermediate ions Q (S$^{2-}$, Se$^{2-}$ and Te$^{2-}$) matters in magnetic interactions

As the radii of the intermediate ions grow, the narrow local space of the magnetic interaction $J$1 becomes overlapped by the Se$^{2-}$ and Te$^{2-}$ ions to a larger extent than by the smaller S$^{2-}$ ions. As a result, a minor displacement of the intermediate ions Se$^{2-}$ and Te$^{2-}$ from the Mo-Mo bond line keeps this line still overlapped and accounts for a contribution to the AFM component of the interaction. As a result, neither MoSe$_2$ nor MoTe$_2$ exhibit an abrupt decrease in the strength of the AFM nearest-neighbor interactions $J$1, nor do the next-nearest-neighbor interactions $J$3($J$1$_2$) undergo FM↔AFM transitions (Supplementary Note 1, Tables 1 and 2). Additionally, the size of the intermediate ions of S, Se and Te has a prominent effect on the relative strength of magnetic interactions. The maximum strengths of the AFM interactions $J$1 are -0.0330 Å$^{-1}$ in

MoS$_2$ (ICSD-76370 [18]), -0.0403 Å$^{-1}$ in MoSe$_2$ (ICSD-644334 [22]) and -0.0615 Å$^{-1}$ in MoTe$_2$ (ICSD-644476 [23]).

In all other aspects, the hexagonal and trigonal structural/magnetic models of MoSe$_2$ and MoTe$_2$ are similar to MoS$_2$ with the dominating AFM nearest-neighbor interactions $J$1. The strong AFM interactions $J$1 form the triangular lattices composed of smaller triangles and compete with each other. The FM next-nearest-neighbor interactions $J$3($J$1$_2$) are so weak ($J$3/$J$1 = -0.11–0.53) that they cannot compete against AFM $J$1. AFM $J$2 ($J$2/$J$1 = 0.06-0.11).

The AFM nearest-neighbor out-of-plane couplings $J$4 are weak ($J$4/$J$1 = 0.16-0.29) in all the dichalcogenides considered. The FM out-of-plane interactions $J$5 are much stronger in the two-layer dichalcogenides 2H-MoSe$_2$ than in their three-layer pars 3R-MoSe2 ($J$5 = 0.0058 Å$^{-1}$, d(Mo-Mo) = 7.499 Å), but weaker than in 2H-MoS$_2$.

## 3.1.5. Exchange magnetic interactions are responsive to displacements of the magnetic ions Mo$^{4+}$

We have previously shown how magnetic interactions undergo transitions following displacements of the intermediate ions of sulfur and selenium n MoS$_2$ and MoSe$_2$ within two centrosymmetric space groups: hexagonal $P$6$_3$/$mmc$ (N194) and trigonal $R3m$ (160). However, the most striking effect can be observed when the magnetic ions Mo$^{4+}$ are displaced in the $ab$ plane, if we compare two MoTe$_2$ polymorphs: hexagonal $P$6$_3$/$mmc$ (N194) (Fig. 5 (a), (b) and (c); Supplementary Note 1 Table 2, Figure 6 (a)-(b)), and high-temperature monoclinic $P$2$_1$/$m$ (N11) (ICSD 14349) [6] (Figure 5 (d –g); Supplementary Note 1 Table 3, Figure 6 (c)-(j)). In the high-temperature modification of MoTe$_2$, the molybdenum atoms lose their central positions in the MoTe$_6$ octahedra and the adjacent rows of the triangle lattice of the magnetic ions of molybdenum become put closer towards each other along the $a$-axis towards in a pairwise manner (Figure 5 (d), (e) and (f)). Due to this displacement, the centrosymmetric hexagonal structure (space group $P$6$_3$/$mmc$) becomes centrosymmetric monoclinic (space group P2$_1$/m (N11)). As a result, the frustrated hexagonal 2D AFM triangular lattice in the $ab$ plane turns into two types of 1D triangular ladders that alternate along the $a$-axis and run along the $b$-axis. Two strong interactions, $J$3 and $J$6, form 1D AFM ladders in the high-temperature polymorph of MoTe$_2$. The third longest AFM interaction $J$3 ($J$3 = -0.0962 Å$^{-1}$, d(Mo1-Mo1) = 4.386 Å) (Fig. 3(d), (e) and (f)) is dominating, and AFM $J$6 ($J$6 = -0.0760 Å$^{-1}$, d(Mo1-Mo1) = 6.581 Å $J$6/$J$3 = 0.78) (Fig. 5(e) and (f)) is only slightly weaker. These interactions are in competition with the weaker AFM interactions $J2$ ($J2 = -0.0272$ Å$^{-1}$, d(Mo1-Mo1) = 3.469 Å $J2/J3 = 0.28$ and $J2/J6 = 0.36$) in the linear sequences along the *b* axis. The legs of the ladder between them is formed by the weaker FM interactions $J1$ ($J1 = 0.0334$ Å$^{-1}$, d(Mo1-Mo1) = 2.893 Å $J1/J3 = -0.34$).

The sublattice of the magnetic ions in the monoclinic MoTe$_2$ appears as Mo1 and Mo2

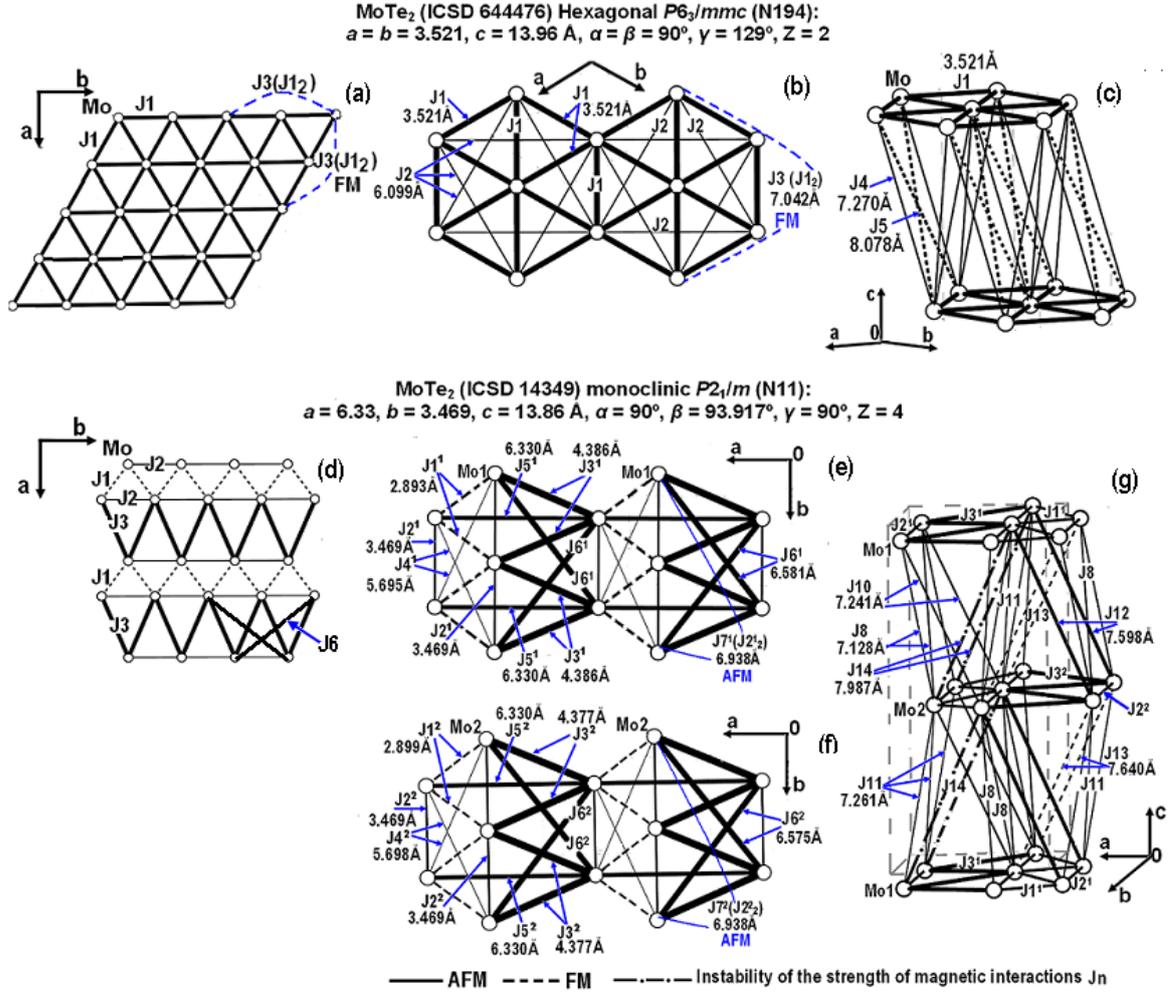

Fig. 5. Displacement of the magnetic ions Mo$^{4+}$ in the ab plane causes major changes in magnetic interactions.

planes alternating each other (Fig. 5 (e) and (f)). The geometry and parameters of the magnetic interactions in the Mo1 and Mo2 planes are virtually identical.

The strength of the nearest-neighbor AFM magnetic interactions $J8$-$J11$ and FM $J13$ between the Mo1 and Mo2 planes (d(Mo1-Mo2) = 7.128 -7.261 Å) is eight times as weak as that of the dominating AFM interaction $J3$ in the planes.

## 3.1.6. Quantitative changes in magnetic interactions turn qualitative as the dimensions of the unit cell decrease or increase

*2H-MoQ$_2$ (Q = S, Se, Te).* Let us see how the characteristics of the main exchange magnetic interactions, AFM *J*1 (*J*1 = -0.0316 Å$^{-1}$, d(Mo1-Mo1) = 3.161 Å) and FM *J*3(*J*1$_2$) (*J*3 = 0.0175 Å$^{-1}$, d(Mo1-Mo1) = 6.322 Å), will change in the *ab* plane of the 2H-MoS$_2$ sample (ICSD 105091 [18]) (Fig. 3) following a compression and an expansion of the original parameters of the unit cell (*a* = *b* = 3.161 Å, *c* = 12.295 Å) without change in symmetry. We have previously demonstrated that the AFM interactions *J*1 in this MoS$_2$ sample are dominating and compete with each other in the smaller triangles. The interactions *J*3(*J*1$_2$) are ferromagnetic, 1.8 times as weak as AFM *J*1 and are not competing with each other in the larger triangles. Additionally, there is no competition between the nearest-neighbor AFM *J*1 and the next-nearest-neighbor FM *J*3(*J*1$_2$) along the -Mo-Mo-Mo- chains in the *ab* plane.

According to our calculations, an in-plane compression following a decrease in the cell parameters along the *a*- and *b*-axis by 0.2 Å results in a 1.6-fold increase in the strength of the AFM interaction *J*1 and a two-fold decrease in the strength of the FM interaction *J*3(*J*1$_2$), while an expansion of these parameters by 0.2 Å causes an opposite effect: a 1.85-fold decrease in the strength of the AFM interaction *J*1 and a 1.4-fold increase in the strength of the FM interaction *J*3(*J*1$_2$).

A compression or an expansion of the cell parameters along the *c*-axis acts in the same direction. A decrease by 0.2 Å leads to a 1.26-fold increase in the strength of the AFM interaction *J*1 and a 1.15-fold decrease in the strength of the FM interaction *J*3(*J*1$_2$). By contrast, an increase by 0.2 Å leads to a 1.36-fold decrease in the strength of the AFM interaction *J*1 and a 1.13-fold increase in the strength of the FM interaction *J*3(*J*1$_2$).

A simultaneous compression of the cell's parameters along the *a*-, *b* -, and *c*-axis by 0.2 Å enhances the above tendencies, and so does the cell's' simultaneous expansion by the same value. Now compression leads to a two-fold increase in the strength of the AFM interaction *J*1 and a three-fold decrease in the strength of the FM interaction *J*3(*J*1$_2$) relative to the original values. By contrast, expansion leads to a three-fold decrease in the strength of the AFM interaction *J*1 and a 1.5-fold increase in the strength of the FM interaction *J*3(*J*1$_2$). Thus, in all the above cases, MoS$_2$ undergoes only quantitative changes in AFM *J*1 and FM *J*3(*J*1$_2$), with no magnetic transitions occurring.

However, a further expansion of the cell parameters by 0.3 Å in the *ab* plane (that is, to $a = b = 3.461$ Å) and by 0.4 Å along the *c*-axis (that is, to $c = 12.695$ Å) leads to AFM↔FM transitions in both interactions, $J1$ and $J3(J1_2)$. In other words, we witness the transitions from quantity to quality. In so doing, the Mo-S distances in the $MoS_6$ octahedra increase by 0.166 Å relative the original value in the 2H-$MoS_2$ sample (ICSD 105091 [18]). These structural changes account for an AFM→FM transition in the interaction $J1$ and an 11-fold decrease in its strength (FM $J1 = 0.0028$ Å$^{-1}$, d(Mo1-Mo1) = 3.461 Å) as well as an FM→AFM transition in the interaction $J3(J1_2)$ and a 1.5 increase in its strength (AFM $J3(J1_2) = -0.0269$ Å$^{-1}$, d(Mo1-Mo1) = 6.922 Å). These transformations of the strong AFM interactions $J1$ into weak FM eliminate frustration in the smaller triangles, but induce frustration of the strong AFM interaction $J3(J1_2)$ in the larger triangles. And yet, despite all the changes in the parameters of the magnetic interactions listed, there is no competition in the -Mo-Mo-Mo- rows, because AFM $J3(J1_2)$ are 9.6 times stronger than FM $J1$.

In the $MoSe_2$ sample (ICSD 644334 [22]), a compression or an expansion of the unit cell parameters by not more than 0.2 Å leads to changes only in the strength of the magnetic interactions, while the direction of the AFM $J1$ и FM $J3(J1_2)$ spins remain unchanged, as with $MoS_2$. A further expansion of the cell parameters by 0.3 Å in the *ab* plane and by 0.4 Å along the *c*-axis (ΔMo-Se = 0.165 Å) leads to a six-fold decrease in the strength of the AFM interaction $J1$ (AFM $J1 = -0.0066$ Å$^{-1}$, d(Mo1-Mo1) = 3.588 Å) in the smaller triangles, as with $MoS_2$; however, the directions of the AFM spins remain the same and the AFM magnetic interactions $J1$ are still in competition. In the larger triangles, the interactions $J3(J1_2)$ undergo an FM→AFM transition, as in $MoS_2$, at much the same values of the strength of the magnetic interactions (AFM $J3(J1_2) = -0.0258$ Å$^{-1}$, d(Mo1-Mo1) = 7.176 Å). Additionally, there is competition running along the -Mo-Mo-Mo- rows between the AFM nearest-neighbor coupling $J1$ and the AFM next-nearest-neighbor coupling $J3(J1_2)$, since the $J1/J3(J1_2)$ ratio, which equals 0.26, exceeds the critical value, no matter whether it is set at 0.16 or 0.24 [26-32]. Thus, expansion of the $MoSe_2$ lattice parameters leads to a stronger competition between magnetic interactions than that of $MoS_2$.

In the $MoTe_2$ sample (ICSD 644476 [23]) (Fig. 5 (a-c)), an expansion of the unit cell parameters by more than 0.2 Å leads to a decrease in the strength of AFM interaction $J1$ and an increase in the strength of FM interaction $J3(J1_2)$. However, in contrast to the $MoS_2$ and $MoSe_2$ situations, the direction of their spins in $MoTe_2$ is unchanged. A sharp 1.4-fold increase in the strength of AFM interaction $J1$, an FM→AFM transition in $J3(J1_2)$ and a 1.81-fold decrease in its strength are observed following a compression of the parameters by 0.2 Å along each of the $a$-, $b$-, and $c$-axis ($\Delta$Mo-Te = 0.103 Å). Now not only the AFM interactions $J1$ in the smaller triangles are frustrated, but also the AFM interactions $J3(J1_2)$ in the larger triangles; however, no frustration is observed to occur along the -Mo-Mo-Mo- rows, because $J3(J1_2)$ are 24 times stronger than AFM $J1$ ($J3(J1_2)/J1=0.041$).

## 3.1.7. The double hydroxides $(M^{2+})_6Al_3(OH)_{18}[Na(H_2O)_6](SO_4)_2 \cdot 6H_2O$ ($M^{2+} = Mn^{2+}$, $Fe^{2+}$).

The double hydroxides $(M^{2+})_6Al_3(OH)_{18}[Na(H_2O)_6](SO_4)_2 \cdot 6H_2O$ ($M^{2+} = Mn^{2+}$, $Fe^{2+}$) [33, 34] may have promise for straintronics. Let us take a closer look at shigaite, a double hydroxide mineral with structural and magnetic parameters being close to those of nikischerite (Supplementary Note 1, Table 4). The shigaite $(Mn^{2+})_6Al_3(OH)_{18}[Na(H_2O)_6](SO_4)_2 \cdot 6H_2O$ [33] crystallizes in the centrosymmetric trigonal/rhombohedral space group R-3 H (N148, ICSD-82492), $a = b = 9.512$, $c = 33.074$ Å, $\alpha = \beta = 90º$, $\gamma = 120º$ Z = 3. No matter how complex these compounds may seem, their magnetic sublattice is primitive. It appears as thin hexagonal layers of $MnO_6$ octahedra (where d(Mn-6O) = 2.172 - 2.209 Å) linked together by shared edges into an open-work hexagonal lattice (Fig. 6 (a)).

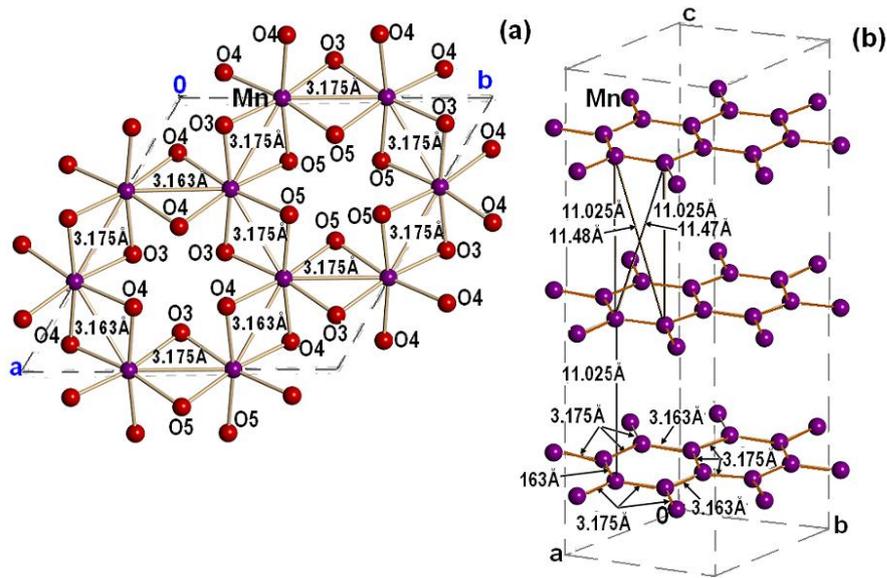

Fig. 6. Hexagonal layers of $MnO_6$ (a) octahedra and the $Mn^{2+}$ sublattice in the hydroxide $(Mn^{2+})_6Al_3(OH)_{18}[Na(H_2O)_6](SO_4)_2 \cdot 6H_2O$.

These magnetic layers are well spaced-out (~11Å) by magnetically neutral oxide layers (Fig. 6 (b)).

The structure of the planes of the magnetic ions $Mn^{2+}$ in the layered double hydroxide $(Mn^{2+})_6Al_3(OH)_{18}[Na(H_2O)_6](SO_4)_2\ 6H_2O$ (Fig. 7) is similar to that of graphene.

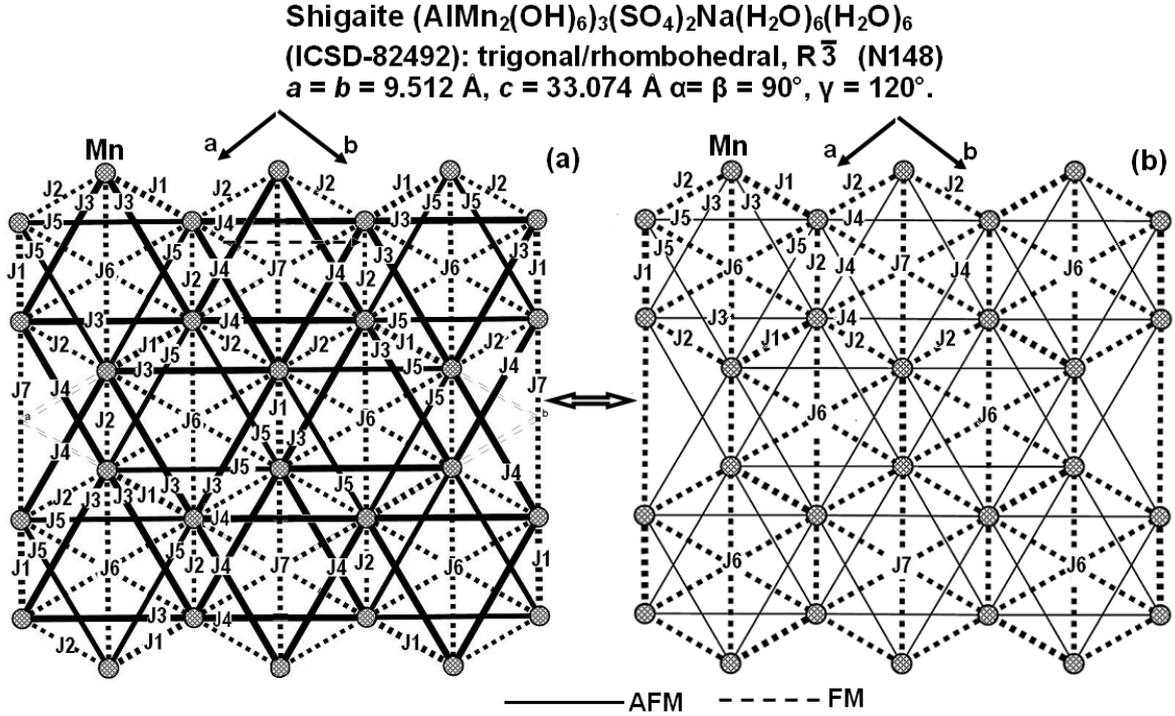

Fig. 7. Interactions $J_n$ in a one-Mn-atom-thick graphene-type layer in shigaite $(AlMn_2(OH)_6)_3(SO_4)_2Na(H_2O)_6(H_2O)_6$. The parameters of magnetic couplings (strength and sign) are presented for initial structural data on shigaite (ICSD-82492) (a) and for a unit cell compressed along each of the a-, b- and c-axis by 0.2 Å (b).

Graphene is ordinarily pictured as a single plane of layered graphite formed by regular carbon hexagons set apart from the bulk crystal. In the double hydroxides, only one third of $Mn^{2+}$ ($Fe^{2+}$) hexagons are regular. This inconsistence has an implication on the magnetic interactions $J_n$.

In the irregular $Mn^{2+}$ hexagons, each of the three sides out of six is d(Mn1-Mn1) = 3.163 Å (we denote the magnetic interaction along these sides by $J1$), and each of the other there d(Mn1-Mn1), is longer, 3.175 Å ($J2$). Six short diagonals form two triangles inscribed in the hexagon. The triangles are equilateral, the sides of one of them being d(Mn1-Mn1) = 5.471 Å (*J*3) each and the sides of the other being d(Mn1-Mn1) = 5.505 Å (*J*5) each. Each long diagonal of the irregular hexagons is d(Mn1-Mn1) = 6.337 Å (*J*6) in length.

In the regular hexagons, each sides is d(Mn1-Mn1) = 3.175 Å (*J*2), each short diagonal is d(Mn1-Mn1) = 5.499 Å (*J*4) and each long diagonal is d(Mn1-Mn1) = 6.349 Å (*J*7). Consequently, the hexagons can be divided into two types: those each having the diagonal formed by *J*6 and a pair of the inscribed triangles *J*3-*J*3-*J*3 and *J*5-*J*5-*J*5, which are not identical, and those having the diagonal formed by *J*7 and a pair of the inscribed triangles *J*4-*J*4-*J*4, which are identical (Fig. 7).

According to our calculations (Supplementary Note 1, Table 4, Fig. 10 (a)-(g)), the strongest interactions are the nearest-neighbor FM *J*1 (*J*1 = -0.0558 Å$^{-1}$, d(Mn1-Mn1) = 3.163 Å) (Figure 10 (a)) along the three short sides of the hexagon. These interactions are 1.4 times stronger than FM *J*2 (*J*2 = -0.0398 Å$^{-1}$, d(Mn1-Mn1) = 3.175 Å, *J*2/*J*1 = 0.71) (Figure 10 (b)) along the three long sides. FM *J*1 are contributed to by two intermediate ions O4; while FM *J*2, by the intermediate ions O3 and O5.

All the interactions along the short diagonals that form triangles are strong AFM *J*3 (*J*3 = -0.0506 Å$^{-1}$, d(Mn1-Mn1) = 5.471 Å, *J*3/*J*1 = -0.91) (Supplementary Fig. 10 (c)), AFM *J*4 (*J*4 = -0.0503 Å$^{-1}$, d(Mn1-Mn1) = 5.499 Å, *J*4/*J*1 = -0.90) (Figure 10 (d)) and AFM *J*5 (*J*5 = -0.0499 Å$^{-1}$, d(Mn1-Mn1) = 5.505 Å, *J*5/*J*1 = -0.89) (Figure 10 (c)). AFM *J*3 and AFM *J*5 are each contributed to by the intermediate ions O3 and O4; while AFM *J*4, by two intermediate ions O5.

However, *J*3, *J*4 and *J*5 are unstable interactions. The reason for the decrease in the strength of the AFM magnetic interactions *J*3, *J*4 and *J*5 and the feasibility of AFM↔FM transitions (which, for example, may happen following compression of the parameters of a cell unit), is the presence of the intermediate ions of manganese close to the boundaries of the local spaces (critical position "a") (Supplementary Note 1, Figure 10 (c)-(l))). A minor displacement of intermediate ions of manganese deeper inside this space is followed by a contribution to the FM component of these interactions so solid that it may even prevail over the contribution to the AFM component of the same interactions. Once these intermediate ions of manganese have crossed the boundary of the local space, they stop contributing to the FM component. We have calculated the changes that are likely to occur to the parameters of the magnetic interactions following a compression of the unit cell of shigaite's crystal structure by as little as 0.2 Å (Figure 9, Supplementary Note 1, Table 4, Figure 10 (c)-(l)).

The interactions along the long diagonals of the hexagons are strong FM *J*6 (*J*6 = 0.0485 Å$^{-1}$, d(Mn1-Mn1) = 6.337 Å, *J*6/*J*1 = 0.87) (Figure 10 (f)) and FM *J*7 (*J*7 = 0.0468 Å$^{-1}$, d(Mn1-Mn1) = 6.349 Å, *J*7/*J*1 = 0.84) (Figure 10 (f)). FM *J*6 are mainly contributed to by the intermediate ions O3 and O4; while FM *J*7, by two intermediate ions O5.

Thus, according to our calculations, all the AFM and FM interactions *J*1-*J*7 in the crystal structure of shigaite are strong. Their minimum ratio is *J*2/*J*1 = 0.71, allowing them to be competing in the triangles that form this structure. First, frustration occurs in the triangles inscribed in the hexagon, when the three parameters are antiferromagnetic at once: AFM*J*3-AFM*J*3-AFM*J*3, AFM*J*4-AFM*J*4-AFM*J*4 and AFM*J*5-AFM*J*5-AFM*J*5. Secondly, here we have yet another type of frustration, when one of the parameters is antiferromagnetic, while the other two are ferromagnetic. In our situation, this type of frustration is specific to triangles AFM*J*3-FM*J*1- FM*J*2, AFM*J*5-FM*J*1- FM*J*2, AFM*J*4-FM*J*2- FM*J*2, AFM*J*3-FM*J*1- FM*J*6, AFM*J*3-FM*J*2- FM*J*6, AFM*J*5-FM*J*1- FM*J*6, AFM*J*5-FM*J*2- FM*J*6 and AFM*J*4-FM*J*2- FM*J*7.

The situation changes dramatically as the dimensions of the unit cell decrease. A 0.2-Å decrease leads to an abrupt 16-fold decrease in AFM *J*3, a 12-fold decrease in AFM *J*4 and a 10-fold decrease in AFM *J*5 (Supplementary Note 1 Table 4, Figure 10 (a)-(g)) while FM *J*1, *J*2, *J*6 and *J*7 remain virtually unaffected. As a result, the competition in the triangles, where the AFM interactions *J*3, *J*4 and *J*5 are weakened but still present, stops, because the ratio of the strengths of the magnetic interactions *J*n (*J*3/*J*1 = -0.07, *J*3/*J*2 =-0.10, *J*3/*J*6 = -0.07, *J*4/*J*2 = -0.13, *J*4/*J*7 = -0.09, *J*5/*J*1 = -0.11, *J*5/*J*2 = -0.16 and *J*5/*J*6 = -0.11) does not exceed the critical value of 0.17 (1/6) [29].

In contrast, a 0.2-Å increase is not even closely as impactful (Supplementary Note 1 Table 4).

The double hydroxides $(M^{2+})_6Al_3(OH)_{18}[Na(H_2O)_6](SO_4)_2$ $6H_2O$ ($M^{2+}$ = $Mn^{2+}$, $Fe^{2+}$) have promise not only because of the transitions that occur following compression, but also because that magnetic interactions between the layers are virtually non-existent. The closest distance between the $Mn^{2+}$ planes is d(Mn1-Mn1) = 11.025 Å and the strength of magnetic interactions between them is as low as FM J$_{interplane}$ = 0.0045 Å$^{-1}$.

## 4. Conclusions

We have considered mechanical deformation as a phenomenon that relates to atomic motions and so demonstrated the feasibility of regulating the magnetic properties of layered van der Waals materials by mechanical strain applied to crystal structures.

We have calculated the parameters of magnetic interactions in the two-dimensional antiferromagnetic dichalcogenides $MoQ_2$ (Q = S, Se, Te) and double hydroxides $(M^{2+})_6Al_3(OH)_{18}[Na(H_2O)_6](SO_4)_2$ $6H_2O$ ($M^{2+}$ = $Mn^{2+}$, $Fe^{2+}$) and developed structural/magnetic models of these compounds. We have established that some of the intermediate ions that have reached the local space of the exchange interaction between magnetic ions take on positions close to critical ones. We have demonstrated that even a minor displacement of these ions following mechanical strain may lead to AFM↔FM transitions or an aberrant change in the strength of these magnetic interactions, no matter whether with or without change in symmetry. This is the main reason why mechanical strain induces magnetic transitions, including transitions from a frustrated to an ordered state.

Thus we have demonstrated that the fluctuations of the intermediate ions near critical positions due to mechanical strain, compression and expansion causing sharp changes to the magnetic parameters allow the magnetic properties to be modified by mechanical strain. To be sure, transition metal dichalcogenides do serve the purpose of being a source of new 2D materials; however, a broad range of transition-metal-based double hydroxides should, in our opinion, be a hefty addition to straintronics' wish list.

### Additional information

Supplementary information is available in the online version of the paper.

### Credit autorshship contributio

**L.M. Volkova:** Writing – original draf

### Declaration of competing interests

The authors declare that they have no known competing financial interests or personal relationships that could have appeared to influence the work reported in this paper.

**Acknowledgments**

The work was financially supported within the frames of the State Order of the Institute of Chemistry FEBRAS, project No. 0205-2023-0001.

**Data availability statement**

All data that support the findings of this study are included within the article (and any supplementary files)..

**ORCID iDs**

LMVolkova https://orcid.org/0000-0002-6316-8586

**Structural-induced magnetic transitions in layred MoQ2 (Q = S, Se, Te) and $(M^{2+})_6Al_3(OH)_{18}[Na(H_2O)_6](SO_4)_2$ $6H_2O$ ($M^{2+} = Mn^{2+}$, $Fe^{2+}$) under the action of mechanical deformations**

*L M Volkova*

**Supplementary Note 1:**

**Figure 5 (all magnetic interactions ($J$n) are shown)**

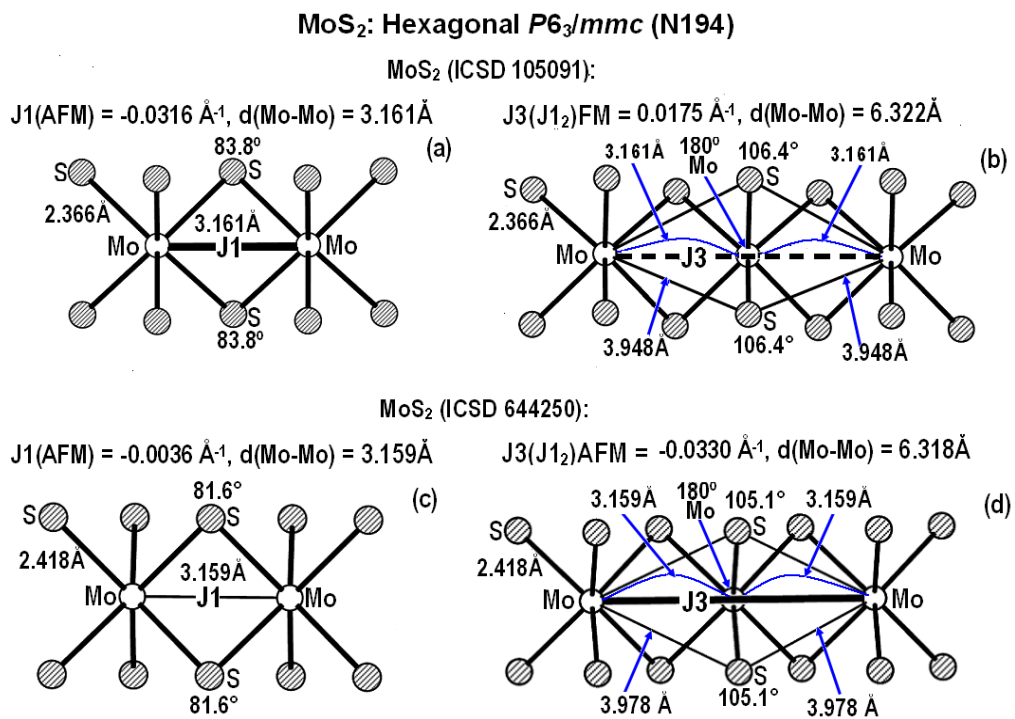

**Figure 5.** The arrangement of intermediate ions in the local spaces of $J1$ and $J3(J1_2)$ в двух образцах MoS$_2$: (ICSD 105091 (a) and (b) and ICSD 644250 (c) and (d)).

**Figure 6 (all magnetic interactions (*J*n) are shown)**

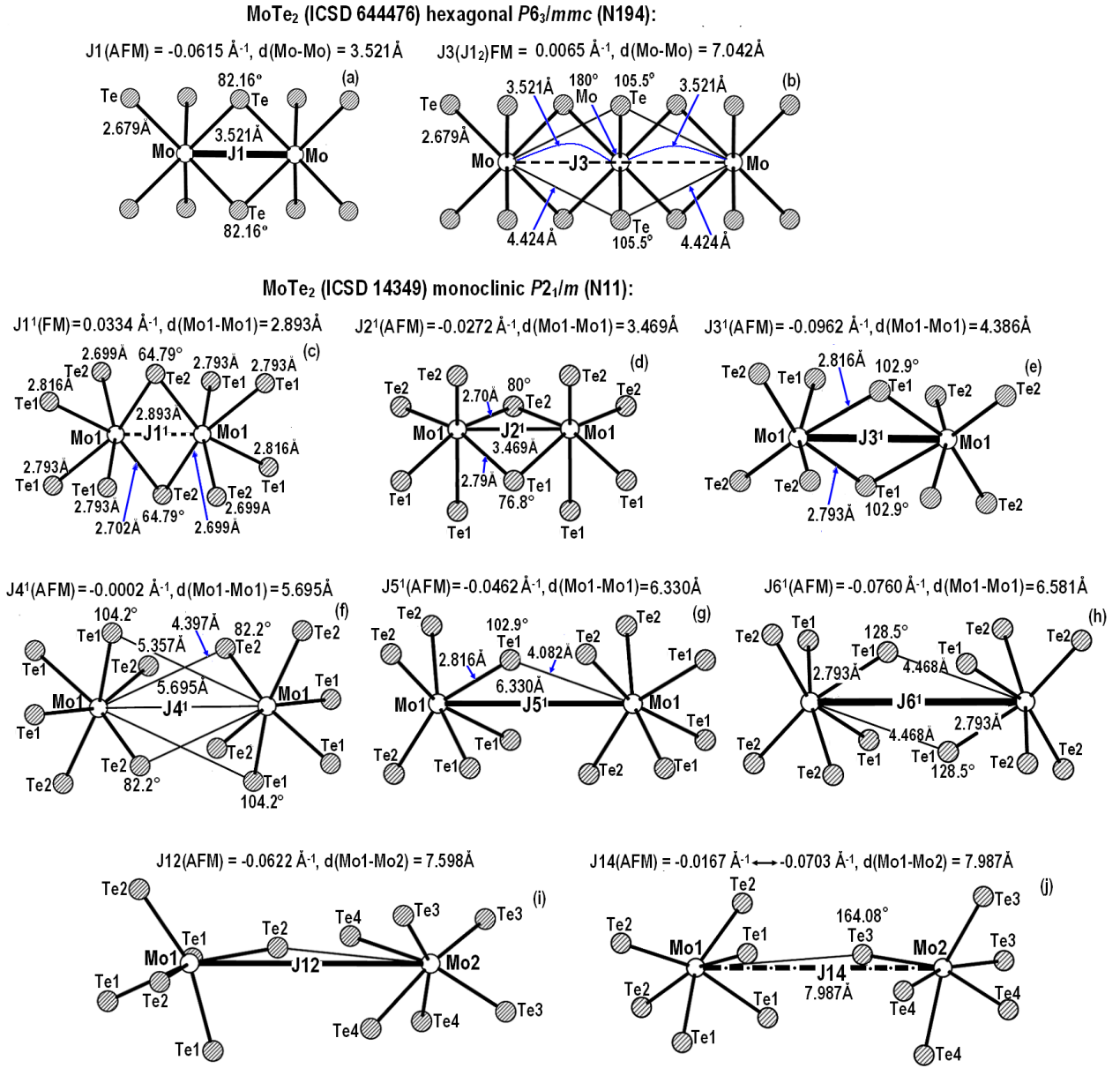

**Figure 6.** The arrangement of intermediate ions in the local spaces of *J*1 (a) and *J*3(*J*1$_2$) (b) in hexagonal *P*6$_3$/*mmc* MoTe$_2$ (ICSD 644476) and in the local spaces of *J*1$^1$ (c), *J*2$^1$ (d), *J*3$^1$ (e), *J*4$^1$ (f), *J*5$^1$ (g), *J*6$^1$ (h), *J*12 (i) and *J*14 (j) in monoclinic *P*2$_1$/*m* MoTe$_2$ (ICSD 14349).

**Figure 10 (all magnetic interactions (*J*n) are shown)**

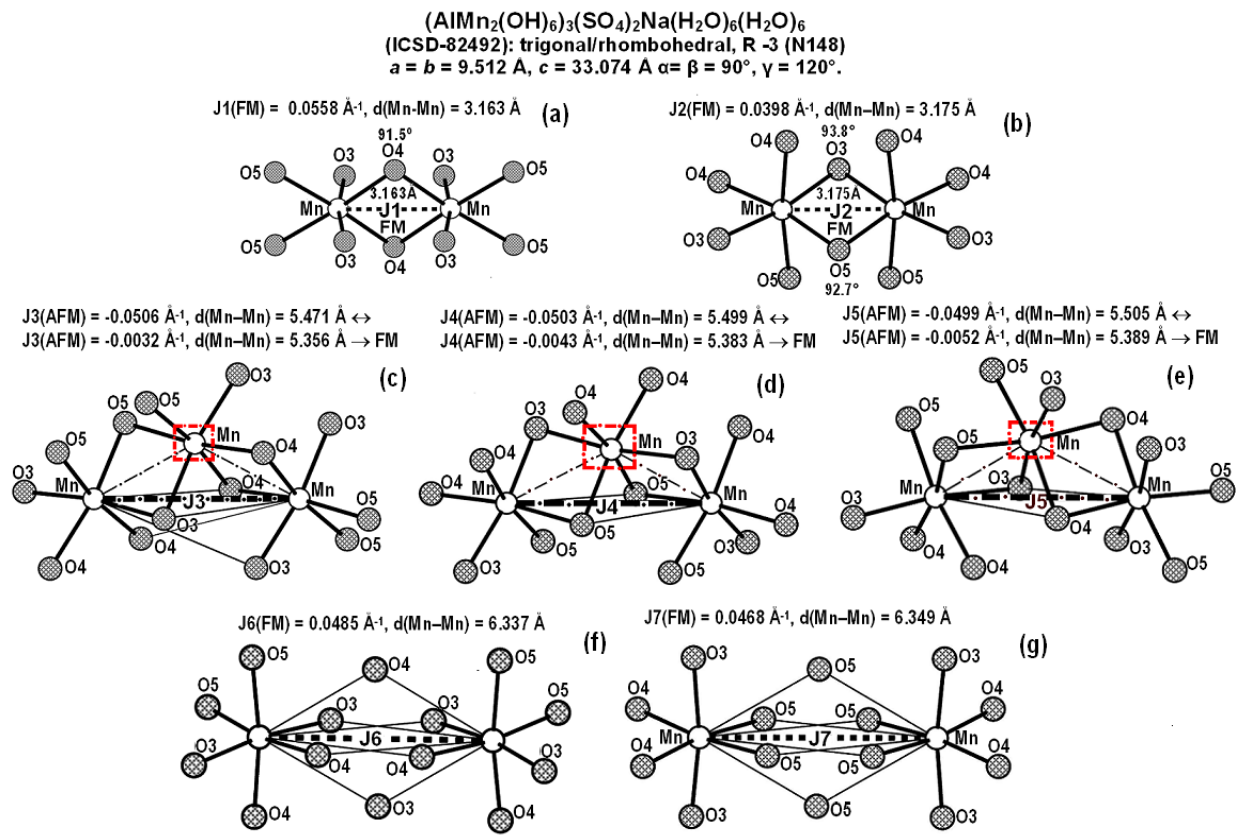

**Figure 10.** The arrangement of intermediate ions in the local spaces of $J1^1$ (a), $J2^1$ (b), $J3^1$ (c), $J4^1$ (d), $J5^1$ (e), $J6^1$ (f) and $J7$ (g) in trigonal/rhombohedral R -3H shigaite $(AlMn_2(OH)_6)_3(SO_4)_2Na(H_2O)_6(H_2O)_6$.

**Table 1. Crystallographic characteristics and parameters of magnetic couplings ($J$n) calculated on the basis of structural data and respective distances between magnetic Mo$^{4+}$ ions in the 2H- and 3R-MoS$_2$**

| Crystallographic and magnetic parameters | MoS$_2$ [18] Min Name: Molybdenite – 2H (Data for ICSD-105091) Space group $P6_3/mmc$ (N194) $a=b=3.161$, $c=12.295$Å $\alpha=\beta=90°$, $\gamma=120°$, Z=2 Method$^{(a)}$: XDP (300K); $R$-value$^{(b)}$ = 0.019 | MoS$_2$ [19] Min Name: Molybdenite – 2H (Data for ICSD-644250) Space group $P6_3/mmc$ (N194) $a=b=3.159$, $c=12.307$Å $\alpha=\beta=90°$, $\gamma=120°$, Z=2 Method$^{(a)}$: XDP (300K); $R$-value$^{(b)}$ = No | MoS$_2$ [18] Min Name: Molybdenite 3R (Data for ICSD - 76370) Space group $R3m$ (N160) $a=b=3.163$, $c=18.37$Å $\alpha=\beta=90°$, $\gamma=120°$, Z=3 Method$^{(a)}$: XDP (293K); $R$-value$^{(b)}$ = 0.026 | MoS$_2$ [20] Min Name: Molybdenite 3R (Data for ICSD - 38401) Space group $R3m$ (N160) $a=b=3.166$, $c=18.410$Å $\alpha=\beta=90°$, $\gamma=120°$, Z=3 Method$^{(a)}$: XDS (293K); $R$-value$^{(b)}$ = 0.086 |
|---|---|---|---|---|
| d(Mo-S) (Å) | Mo: trigonal prism Mo – S1 = 2.366x6 | Mo: trigonal prism Mo – S1 = 2.418x6 | Mo: trigonal prism Mo – S1 = 2.364x3 – S2 = 2.365x3 | Mo: trigonal prism Mo – S1 = 2.414x3 – S2 = 2.414x3 |
| *Triangle plane* | Hexagonal Mo 2c: 0.333 0.667 0.25 S1 4f: 0.333 0.667 0.627 | Hexagonal Mo 2c: 0.333 0.667 0.25 S1 4f: 0.333 0.667 0.621 | Rhombohedral Mo 3a: 0 0 0 S1 3a: 0 0 0.252 S1 3a: 0 0 0.415 | Rhombohedral Mo 3a: 0 0 0 S1 3a: 0 0 0.248 S1 3a: 0 0 0.419 |
| d(Mo-Mo) (Å) | 3.161 | 3.159 | 3.163 | 3.166 |
| $J1^{(c)}$ (Å$^{-1}$) | $J1$ = -0.0316 (AFM) | $J1$ = -0.0036 (AFM) | $J1$ = -0.0330 (AFM) | $J1$ = -0.0070 (AFM) |

| $j(X)^d$ (Å$^{-1}$) | $j$(S1): -0.0158x2 | $j$(S1): -0.0018x2 | $j$(S1): -0.0165 | $j$(S1): -0.0035x1 |
|---|---|---|---|---|
| ($\Delta h(X)^e$ Å, $l_n'/l_n^f$, MoXMo$^g$) | (-0.079 1.0, 83.81°) | (-0.009 1.0, 81.57°) | (-0.083 1.0, 83.97°) | (-0.018 1.0, 81.96°) |
| $j(X)^d$ (Å$^{-1}$) | $J1/J1 = 1$ | $J1/J3 = 0.11$ | $j$(S2): -0.0164 | $j$(S2): -0.0034x1 |
| ($\Delta h(X)^e$ Å, $l_n'/l_n^f$, MoXMo$^g$) | | | (-0.082 1.0, 83.96°) | (-0.017 1.0, 81.94°) |
| $Jn/J$max | | | $J1/J1 = 1$ | $J1/J3 = 0.21$ |
| | | | | |
| d(Mo-Mo) (Å) | 5.475 | 5.472 | 5.478 | 5.484 |
| $J2^{(c)}$ (Å$^{-1}$) | $J2^* = -0.0020$ (AFM) | $J2^* = 0.0024$ (FM) | $J2^* = -0.0020$ (AFM) | $J2^* = 0.0016$ (FM) |
| $j(X)^d$ (Å$^{-1}$) | $j$(S1): -0.0056x2 | $j$(S1): -0.0042x2 | $j$(S1): -0.0056 | $j$(S1): -0.0044 |
| ($\Delta h(X)^e$ Å, $l_n'/l_n^f$, MoXMo$^g$) | (-0.334, 2.0, 118.04°) | (-0.252, 2.0, 115.44) | (-0.339, 2.0, 118.23°) | (-0.263, 2.0, 115.88°) |
| $j(X)^d$ (Å$^{-1}$) | $j$(S1): 0.0023x4 | $j$(S1): 0.0027x4 | $j$(S2): -0.0056 | $j$(S2): -0.0044 |
| ($\Delta h(X)^e$ Å, $l_n'/l_n^f$, MoXMo$^o$) | (0.343, 5.0, 87.12°) | (0.399, 5.0, 86.0°) | (0.338, 2.0, 118.21°) | (-0.263, 2.0, 115.88°) |
| $j(X)^d$ (Å$^{-1}$) | $J2/J1 = 0.06$ | $J2/J3 = -0.07$ | $j$(S1): 0.0023x2 | $j$(S1): 0.0026x2 |
| $j(X)^d$ (Å$^{-1}$) | | | $j$(S2): 0.0023x2 | $j$(S2): 0.0026x2 |
| $Jn/J$max | | | $J2/J1 = 0.07$ | $J2/J3 = -0.05$ |
| | | | | |
| d(Mo-Mo) (Å) | 6.322 | 6.318 | 6.326 | 6.332 |
| $J3^{(c)}$ (Å$^{-1}$) | $J3^* = 0.0175$ (FM) | $J3^* = -0.0330$ (AFM) | $J3^* = 0.0172$ (FM) | $J3^* = -0.0327$ (AFM) |
| $j(X)^d$ (Å$^{-1}$) | $j$(Mo1): -0.0325 | $j$(Mo1): -0.0330 | $j$(Mo1): -0.0325 | $j$(Mo1): -0.0324 |
| ($\Delta h(X)^e$ Å, $l_n'/l_n^f$, MoXMo$^g$) | (-0.650, 1.0, 180°) | (-0.650, 1.0, 180°) | (-0.650, 1.0, 180°) | (-0.650, 1.0, 180°) |
| $j(X)^d$ (Å$^{-1}$) | $j$(S1): 0.0263x2 | $j$(S1): 0.0290x2 | $j$(S1): 0.0262 | $j$(S1): 0.0286 |
| ($\Delta h(X)^e$ Å, $l_n'/l_n^f$, MoXMo$^g$) | (0.526, 1.0, 106.37°) | (0.578, 1.0, 105.14°) | (0.524, 1.0, 106.45°) | (0.574, 1.0, 106.45°) |
| $j(X)^d$ (Å$^{-1}$) | $J3/J1 = -0.55$ | $J3/J3 = 1$ | $j$(S2): 0.0262 | $j$(S2): 0.0286 |

| ($\Delta h$(X)[e] Å, $l_n$'/$l_n$[f], MoXMo[g]) | | | (0.525, 1.0, 106.44°) | (0.574, 1.0, 106.44°) |
|---|---|---|---|---|
| $J$n/$J$max | | | $J3/J1 = -0.52$ | $J3/J3 = 1$ |
| *Interplane couplings* | | | | |
| d(Mo-Mo) (Å) | 6.413 | 6.418 | 6.390 | 6.403 |
| $J4$[(c)] (Å$^{-1}$) | $J4^* = -0.0064$ (AFM)[h] | $J4^* = -0.0070$ (AFM)[h] | $J4^* = -0.0054$ (AFM)[h] | $J4^* = -0.0056$ (AFM)[h] |
| $j$(X)[d] (Å$^{-1}$) | $j$(S1): -0.0056x2 | $j$(S1): -0.0061x2 | $j$(S1): -0.0056 | $j$(S1): -0.0061 |
| ($\Delta h$(X)[e] Å, $l_n$'/$l_n$[f], MoXMo[g]) | (-0.519, 2.27, 129.53°) | (-0.543, 2.15, 131.04°) | (-0.519, 2.25, 129.43°) | (-0.538, 2.15, 130.78°) |
| $j$(X)[d] (Å$^{-1}$) | $j$(S1): 0.0012x4 | $j$(S1): 0.0013x4 | $j$(S2): -0.0017x2 | $j$(S2): -0.0019x2 |
| ($\Delta h$(X)[e] Å, $l_n$'/$l_n$[f], MoXMo[g]) | (0.209, 4.42, 98.63°) | (0.222, 4.08, 99.68°) | (-0.197, 2.76, 116.67°) | (-0.201, 2.62, 117.74°) |
| $J$n/$J$max | $J4/J1 = 0.20$ | $J4/J3 = 0.21$ | $J4/J1 = 0.16$ | $J4/J3 = 0.17$ |
| | | | | |
| d(Mo-Mo) (Å) | 7.150 | 7.153 | 7.130 | 7.130 |
| $J5$[(c)] (Å$^{-1}$) | *$J5$ = 0.0358 (FM) | *$J5$ = 0.0319 (FM) | *$J5$ = 0.0086 (FM) | *$J5$ = 0.0063 (FM) |
| $j$(X)[d] (Å$^{-1}$) | $j$(S1): 0.0213x2 | $j$(S1): 0.0195x2 | $j$(S1): 0.0213 | $j$(S1): 0.0197 |
| ($\Delta h$(X)[e] Å, $l_n$'/$l_n$[f], MoXMo[g]) | (0.530, 1.26, 112.42°) | (0.488, 1.22, 113.52°) | (0.528, 1.26, 112.35°) | (0.494, 1.21, 113.33°) |
| $j$(X)[d] (Å$^{-1}$) | $j$(S1): -0.0017x4 | $j$(S1): -0.0018x4 | $j$(S2): -0.0093 | $j$(S2): -0.0100 |
| ($\Delta h$(X)[e] Å, $l_n$'/$l_n$[f], MoXMo[g]) | (-0.259, 3.06 121.74) | (-0.260, 2.91 122.68) | (1.041, 2.20 151.0) | (1.077, 2.12, 152.6) |
| $j$(X)[d] (Å$^{-1}$) | $J5/J1 = -1.130$ | $J5/J3 = -0.97$ | $j$(S2): -0.0017x2 | $j$(S2): -0.0017x2 |
| ($\Delta h$(X)[e] Å, $l_n$'/$l_n$[f], MoXMo[g]) | | | (-0.0258, 3.06 121.61) | (-0.0257, 2.83 122.45) |
| $J$n/$J$max | | | $J5/J1 = -0.32$ | $J5/J3 = -0.19$ |
| | | | | |
| d(Mo-Mo) (Å) | 7.817 | 7.820 | 7.800 | 7.813 |
| $J6$[(c)] (Å$^{-1}$) | $J6 = -0.0116$ (AFM) | $J6 = -0.0139$ (AFM) | $J6 = 0.0004$ (FM) | $J6 = -0.0012$ (FM) |

| $j(X)^d$ (Å$^{-1}$) | $j$(S1): -0.0073x2 | $j$(S1): -0.0078x2 | $j$(S1): -0.0074 | $j$(S1): -0.0078 |
|---|---|---|---|---|
| ($\Delta h(X)^e$ Å, $l_n'/l_n^f$, MoXMo$^g$) | (-1.106 2.48, 154.43°) | (-1.134 2.38, 155.72°) | (-1.107 2.47, 154.41°) | (-1.131 2.39, 155.57°) |
| $j(X)^d$ (Å$^{-1}$) | $j$(S1): 0.0007x2 | $j$(S1): 0.0009x2 | $j$(S2): -0.0037 | $j$(S2): -0.0038 |
| ($\Delta h(X)^e$ Å, $l_n'/l_n^f$, MoXMo$^g$) | (0.318 7.04, 96.75°) | (0.344 6.55, 97.53°) | (-0.635 2.83, 137.55°) | (-0.638 2.73, 138.29°) |
| $j(X)^d$ (Å$^{-1}$) | J6/J1 = 0.37 | J6/J3 = 0.42 | $j$(S2): 0.0104 | $j$(S2): 0.0092 |
| ($\Delta h(X)^e$ Å, $l_n'/l_n^f$, MoXMo$^g$) | | | (0.303 1.305, 121.58°) | (0.269 1.31, 122.58°) |
| Jn/Jmax | | | J6/J1 = -0.01 | J6/J3 = 0.037 |
| | | | | |
| d(Mo-Mo) (Å) | 8.363 | 8.358 | 8.369 | 8.376 |
| $J7^{(c)}$ (Å$^{-1}$) | J7* = 0.0136 (FM) | J7* = 0.0189 (FM) | J7* = 0.0134 (FM) | J7* = 0.0185 (FM) |
| $j(X)^d$ (Å$^{-1}$) | $j$(Mo1): 0.0130x2 | $j$(Mo1): 0.0130x2 | $j$(Mo1): 0.0130x2 | $j$(Mo1): 0.0130x2 |
| ($\Delta h(X)^e$ Å, $l_n'/l_n^f$, MoXMo$^g$) | (0.385, 1.80, 150.0°) | (0.384, 1.80, 150.0°) | (0.385, 1.80, 150.0°) | (0.386, 1.80, 150.0°) |
| $j(X)^d$ (Å$^{-1}$) | $j$(S1): -0.0055x2 | $j$(S1): -0.0033x2 | $j$(S2): -0.0056 | $j$(S2): -0.0035 |
| ($\Delta h(X)^e$ Å, $l_n'/l_n^f$, MoXMo$^g$) | (-0.183, 1.33, 136.08°) | (-0.183, 1.33, 134.29°) | (-0.187, 1.33, 136.19°) | (-0.119, 1.33, 134.6°) |
| $j(X)^d$ (Å$^{-1}$) | J7/J1 = -0.43 | J7/J3 = -0.57 | $j$(S1): -0.0056 | $j$(S1): -0.0035 |
| | | | | |
| ($\Delta h(X)^e$ Å, $l_n'/l_n^f$, MoXMo$^g$) | 9.005 | 9.006 | (-0.188, 1.33, 136.20°) | (-0.119, 1.33, 134.6°) |
| Jn/Jmax | J8 = -0.0168 AFM | J8 = -0.0192 AFM | J7/J1 = -0.41 | J7/J3 = -0.57 |
| d(Mo-Mo) (Å) | $j$(S1): -0.0069x2 | $j$(S1): -0.0081x2 | 8.992 | 9.005 |
| $J8^{(c)}$ (Å$^{-1}$) | (-0.257 1.49, 139.98°) | (-0.305 1.45, 141.26°) | J8 = -0.0051 AFM | J8 = -0.0074 AFM |
| | $j$(S1): -0.0017x2 | $j$(S1): -0.0017x2 | J8/J1 = 0.15 | J8/J3 = 0.23 |
| J8/Jmax | (-0.504 3.61, 134.90°) | (-0.493 3.48, 135.26°) | | |
| | J8/J1 = 0.53 | J8/J3 = 0.58 | | |

| d(Mo-Mo) (Å) | 9.483 | 9.477 | 9.489 | 9.498 |
|---|---|---|---|---|
| $J9^{(c)}$ (Å$^{-1}$) | J9 = 0.0003 FM | J9 = -0.0076 AFM | J9 = 0.0177 FM | J9 = 0.0046 FM |
| $J9/J$max | $J9/J1$ = -0.009 | $J9/J1$ = 0.23 | $J9/J1$ = -0.54 | $J9/J3$ = -0.14 |
| | | | | |
| d(Mo-Mo) (Å) | 9.544 | 9.544 | 9.532 | 9.546 |
| $J10^{(c)}$ (Å$^{-1}$) | J10 = 0.0351 FM | J10 = 0.0306 FM | J10 = 0.0011 FM | J10 = -0.0017 AFM |
| $J10/J$max | $J10/J1$ = -1.11 | $J10/J1$ = -0.93 | $J10/J1$ = -0.03 | $J10/J3$ = 0.05 |

[a]XDP: X-ray diffraction from a powder crystal, XDS: X-ray diffraction from a single crystal.
[b]The refinement converged to the residual factor ($R$) values.
[c]$Jn$ in Å$^{-1}$: the magnetic couplings ($Jn < 0$, AFM; $Jn > 0$, FM)
[d]$j(X)$: contributions of the intermediate ion X to the AFM ($j(X) <0$) and FM ($j(X)>0$) components of the coupling $Jn$.
[e]$\Delta h(X)$: the degree of overlapping of the local space between magnetic ions by the intermediate ion X.
[f]$l_n'/l_n$: the asymmetry of the position of the intermediate ion X relative to the middle of the Mo$_i$–Mo$_j$ bond line.
[g]M$_i$XM$_j$: bonding angle.
[h]Small $j(X)$ contributions are not shown.

**Table 2.** Crystallographic characteristics and parameters of magnetic couplings ($J_n$) calculated on the basis of structural data and respective distances between magnetic $Mo^{4+}$ ions in hexagonal 2H-$MoSe_2$ and 2H-$MoTe_2$ and rhombohedral high-pressure high-temperature polymorph 3R-$MoSe_2$

| Crystallographic and magnetic parameters | $MoSe_2$ (2H) [22] ICSD-644334 Space group $P6_3/mmc$ (N194) $a=b=3.288, c=12.900$Å $\alpha=\beta=90°, \gamma=120°, Z=2$ Method[a]: XDP (293K); $R$-value[b] = No | $MoSe_2$ (2H) [24] (ICSD-191306) Space group $P6_3/mmc$ (N194) $a=b=3.289, c=12.927$Å $\alpha=\beta=90°, \gamma=120°, Z=2$ Method[a]: XDP (300K); $R$-value[b] = No | $MoTe_2$ (2H) [23] (ICSD-644476) Space group $P6_3/mmc$ (N194) $a=b=3.521, c=13.96$Å $\alpha=\beta=90°, \gamma=120°, Z=2$ Method[a]: XDS (293K); $R$-value[b] = No | $MoSe_2$ (3R) [21] (ICSD-16948) Space group $R3m$ (N160) $a=b=3.292, c=19.392$Å $\alpha=\beta=90°, \gamma=120°, Z=3$ Method[a]: XDP (293K); $R$-value[b] = No |
|---|---|---|---|---|
| d(Mo-S) (Å) | Mo: octahedron Mo – Se1 = 2.491x6 | Mo: octahedron Mo – Se1 = 2.510x6 | Mo: octahedron Mo – Te1 = 2.679x6 | Mo: octahedron Mo – Se1 = 2.491x3 – Se2 = 2.495x3 |
| *Triangle plane* | Hexagonal Mo 2c: 0.3333 0.6667 0.25 Se1 4f: 0.3333 0.6667 0.625 | Hexagonal Mo 2c: 0.3333 0.6667 0.25 Se1 4f: 0.3333 0.6667 0.623 | Hexagonal Mo 2c: 0.3333 0.6667 0.25 S1 4f: 0.3333 0.6667 0.625 | Trional-Нет центра Mo 3a:0.3333 0.6667 0 Se1 3a:0.6667 0.3333 0.083 Se2 3a: 0.3333 0.6667 0.25 |
| d(Mo-Mo) (Å) | 3.288 | 3.289 | 3.521 | 3.292 |
| $J1$[c] (Å$^{-1}$) | $J1$ = -0.0403 (AFM) | $J1$ = -0.0308 (AFM) | $J1$ = -0.0615 (AFM) | $J1$ = -0.0399 (AFM) |
| $j(X)$[d] (Å$^{-1}$) | $j$(Se1): -0.0202x2 | $j$(Se1): -0.0154x2 | $j$(Te1): -0.0307 x2 | $j$(Se1): -0.0205 |
| ($\Delta h(X)$[e]Å, $l_n'/l_n$[f], MoXMo[g]) | (-0.109 1.0, 82.61°) | (-0.083 1.0, 81.86°) | (-0.191 1.0, 82.16°) | (-0.111 1.0, 82.74°) |
| $j(X)$[d] (Å$^{-1}$) | $J1/J1$ = 1 | $J1/J1$ = 1 | $J1/J1$ = 1 | $j$(Se2): -0.0194 |
| ($\Delta h(X)^e$ Å, $l_n'/l_n^f$, MoXMo$^g$) | | | | (-0.105 1.0, 82.57°) |
| $J$n/$J$max | | | | $J1/J1 = 1$ |
| | | | | |
| d(Mo-Mo) (Å) | 5.695 | 5.697 | 6.099 | 5.702 |
| $J2^{(c)}$ (Å$^{-1}$) | $J2^* = -0.0034$ (AFM) | $J2^* = -0.0019$ (AFM) | $J2^* = -0.0070$ (AFM) | $J2^* = -0.0033$ (AFM) |
| $j(X)^d$ (Å$^{-1}$) | $j$(Se1): -0.0057x2 | $j$(Se1): -0.0052x2 | $j$(Te1): -0.0063x2 | $j$(Se1): -0.0057 |
| ($\Delta h(X)^e$ Å, $l_n'/l_n^f$, MoXMo$^g$) | (-0.367, 2.0, 116.64°) | (-0.338, 2.0, 115.77°) | (-0.465, 2.0, 116.13°) | (-0.370, 2.0, 116.80°) |
| $j(X)^d$ (Å$^{-1}$) | $j$(Se1): 0.0020x4 | $j$(Se1): 0.0021x4 | $j$(Te1): 0.0014x4 | $j$(Se2): -0.0056 |
| ($\Delta h(X)^e$ Å, $l_n'/l_n^f$, MoXMo$^o$) | $J2/J1 = 0.08$ | $J2/J1 = 0.06$ | $J2/J1 = 0.11$ | (0.364, 2.0, 116.60°) |
| $j(X)^d$ (Å$^{-1}$) | | | | $j$(Se1): 0.0020x2 |
| $j(X)^d$ (Å$^{-1}$) | | | | $j$(Se2): 0.0020x2 |
| $J$n/$J$max | | | | $J2/J1 = 0.08$ |
| | | | | |
| d(Mo-Mo) (Å) | 6.576 | 6.578 | 7.042 | 6.584 |
| $J3^{(c)}$ (Å$^{-1}$) | $J3^* = 0.0139$ (FM) | $J3^* = 0.0164$ (FM) | $J3^* = 0.0065$ (FM) | $J3^* = 0.0140$ (FM) |
| $j(X)^d$ (Å$^{-1}$) | $j$(Mo1): -0.0301 | $j$(Mo1): -0.0300 | $j$(Mo1): -0.0262 | $j$(Mo1): -0.0300 |
| ($\Delta h(X)^e$ Å, $l_n'/l_n^f$, MoXMo$^g$) | (-0.650, 1.0, 180°) | (-0.650, 1.0, 180°) | (-0.650, 1.0, 180°) | (-0.650, 1.0, 180°) |
| $j(X)^d$ (Å$^{-1}$) | $j$(Se1): 0.0236x2 | $j$(Se1): 0.0245x2 | $j$(Te1): 0.0189x2 | $j$(Se1): 0.0236 |
| ($\Delta h(X)^e$ Å, $l_n'/l_n^f$, MoXMo$^g$) | (0.511, 1.0, 105.70°) | (0.530, 1.0, 105.39°) | (0.469, 1.0, 105.46°) | (0.511, 1.0, 105.77°) |
| $j(X)^d$ (Å$^{-1}$) | $J3/J1 = -0.34$ | $J3/J1 = -0.53$ | $J3/J1 = -0.11$ | $j$(Se2): 0.0237 |
| ($\Delta h(X)^e$ Å, $l_n'/l_n^f$, MoXMo$^g$) | | | | (0.515, 1.0, 105.69°) |
| $J$n/$J$max | | | | $J3/J1 = -0.35$ |
| *Interplane couplings* | | | | |

| d(Mo-Mo) (Å) | 6.723 | 6.737 | 7.270 | 6.738 |
|---|---|---|---|---|
| $J4^{(c)}$ (Å$^{-1}$) | $J4^* = -0.0089$ (AFM) [h] | $J4^* = -0.0090$ (AFM) [h] | $J4^* = -0.0112$ (AFM) [h] | $J4^* = -0.0076$ (AFM) [h] |
| $j(X)^d$ (Å$^{-1}$) | $j$(Se1): -0.0061x2 | $j$(Se1): -0.0062x2 | $j$(Te1): -0.0063x2 | $j$(Se2): -0.0061 |
| ($\Delta h(X)^e$ Å, $l_n'/l_n^f$, MoXMo$^g$) | (-0.614, 2.23, 130.35°) | (-0.621, 2.19, 130.85°) | (-0.746, 2.23, 130.64°) | (-0.612, 2.23, 130.37°) |
| $j(X)^d$ (Å$^{-1}$) | $j$(Se1): 0.0008x4 | $j$(Se1): 0.0009x4 | $j$(Te1): 0.0004x4 | $j$(Se1): -0.0022x2 |
| ($\Delta h(X)^e$ Å, $l_n'/l_n^f$, MoXMo$^g$) | (0.157, 4.25, 99.46°) | (0.163, 4.15, 99.65°) | (0.079, 4.22, 99.65°) | (-0.271, 2.71, 117.56°) |
| $Jn/J$max | $J4/J1 = 0.22$ | $J4/J1 = 0.29$ | $J4/J1 = 0.18$ | $J4/J1 = 0.19$ |
| | | | | |
| d(Mo-Mo) (Å) | 7.484 | 7.497 | 8.078 | 7.499 |
| $J5^{(c)}$ (Å$^{-1}$) | $*J5 = 0.0268$ (FM) | $*J5 = 0.0256$ (FM) | $*J5 = 0.0177$ (FM) | $*J5 = 0.0058$ (FM) |
| $j(X)^d$ (Å$^{-1}$) | $j$(Se1): 0.0174x2 | $j$(Se1): 0.0169x2 | $j$(Te1): 0.0134x2 | $j$(Se2): 0.0174 |
| ($\Delta h(X)^e$ Å, $l_n'/l_n^f$, MoXMo$^g$) | (0.474, 1.26, 113.52°) | (0.463, 1.24, 113.38°) | (0.425, 1.27, 113.23°) | (0.478, 1.26, 113.03°) |
| $j(X)^d$ (Å$^{-1}$) | $j$(Se1): -0.0018x4 | $j$(Se1): -0.0020x4 | $j$(Te1): -0.0023x4 | $j$(Se1): -0.0094 |
| ($\Delta h(X)^e$ Å, $l_n'/l_n^f$, MoXMo$^g$) | (-0.260, 3.0 122.37) | (-0.335, 2.95 122.71) | (-0.449, 3.0 122.72) | (-1.157, 2.19 151.63) |
| $j(X)^d$ (Å$^{-1}$) | $J5/J1 = -0.66$ | $J5/J1 = -0.83$ | $J5/J1 = -0.29$ | $j$(Se2): -0.0020x2 |
| ($\Delta h(X)^e$ Å, $l_n'/l_n^f$, MoXMo$^g$) | | | | (-0.0334, 3.00 122.40) |
| $Jn/J$max | | | | $J5/J1 = -0.14$ |

[a]XDP: X-ray diffraction from a powder crystal, XDS: X-ray diffraction from a single crystal.
[b]The refinement converged to the residual factor ($R$) values.
[c]$Jn$ in Å$^{-1}$: the magnetic couplings ($Jn < 0$, AFM; $Jn > 0$, FM)
[d]$j(X)$: contributions of the intermediate ion X to the AFM ($j(X) <0$) and FM ($j(X)>0$) components of the coupling $Jn$.
[e]$\Delta h(X)$: the degree of overlapping of the local space between magnetic ions by the intermediate ion X.
[f]$l_n'/l_n$: the asymmetry of the position of the intermediate ion X relative to the middle of the Mo$_i$–Mo$_j$ bond line.
[g]M$_i$XM$_j$: bonding angle.
[h]Small j(X) contributions are not shown.

**Table 3. Crystallographic characteristics and parameters of magnetic couplings (*J*n) calculated on the basis of structural data and respective distances between magnetic Mo$^{4+}$ ions in the high-temperature monoclinic MoTe$_2$.**

| | **MoTe$_2$ ( ICSD-14349) [25]** | |
|---|---|---|
| | Space group *P*2$_1$/*m* (N11) | |
| | *a*=6.33, *b*=3.469, *c*=13.86Å, *α*=90º, *β*=93.917º, *γ*=90º, Z=4 | |
| | Method[a]: XDP; *R*-value[b] = 0.139 | |
| Crystallographic and magnetic parameters | Mo1 layer<br>Mo1: octahedron<br>Mo1 – Te2 = 2.699 Å x2<br>– Te2 = 2.702 Å x1<br>– Te1 = 2.793 Å x2<br>– Te1 = 2.816 Å x1 | Mo2 layer<br>Mo2: octahedron<br>Mo2 – Te3 = 2.699 Å x2<br>– Te3 = 2.710 Å x1<br>– Te4 = 2.789 Å x2<br>– Te4 = 2.810 Å x1 |
| d(Mo-Mo) (Å) | 2.893 | 2.899 |
| *J*1[c] (Å$^{-1}$) | *J*1$^1$ = -0.0334 (FM) | *J*1$^2$ = -0.0348 (FM) |
| *j*(X)$^d$ (Å$^{-1}$) | *j*(Te2): 0.0167x2 | *j*(Te3): 0.0174x2 |
| (Δ*h*(X)$^e$Å, *l*$_n$'/*l*$_n^f$, MoXMo$^g$) | (0.070 1.0, 64.79°) | (0.073 1.0, 64.81°) |
| *J*n/*J*max (*J*max = *J*3$^2$) | *J*1$^1$/*J*3$^2$ = -0.34 | *J*1$^2$/*J*3$^2$ = -0.36 |
| | | |
| d(Mo-Mo) (Å) | 3.469 | 3.469 |
| *J*2[c] (Å$^{-1}$) | *J*2$^1$ = -0.0272 (AFM) | *J*2$^2$ = -0.0279 (AFM) |
| *j*(X)$^d$ (Å$^{-1}$) | *j*(Te2): -0.0237 | *j*(Te3): -0.0236 |
| (Δ*h*(X)$^e$Å, *l*$_n$'/*l*$_n^f$, MoXMo$^g$) | (-0.0143, 1.0, 80.0°) | (-0.0142, 1.0, 80.0°) |
| *j*(X)$^d$ (Å$^{-1}$) | *j*(Te1):- 0.0035 | *j*(Te4):- 0.0044 |
| (Δ*h*(X)$^e$Å, *l*$_n$'/*l*$_n^f$, MoXMo°) | (-0.021, 1.0, 76.8°) | (-0.026, 1.0, 76.9°) |

| $J$n/$J$max | $J2^1/J3^2 = 0.28$ | $J2^2/J3^2 = 0.29$ |
|---|---|---|
| | | |
| d(Mo-Mo) (Å) | 4.386 | 4.377 |
| $J3^{(c)}$ (Å$^{-1}$) | $J3^1$= -0.0962 AFM | $J3^2$= -0.0970 AFM |
| $j(X)^d$ (Å$^{-1}$) | $j$(Te1): -0.0481x2 | $j$(Te4): -0.0485x2 |
| ($\Delta h(X)^e$Å, $l_n'/l_n^f$, MoXMo$^g$) | (-0.462, 1.01, 102.9°) | (-0.464, 1.01, 102.9°) |
| $J$n/$J$max | $J3^1/J3^2 = 0.99$ | $J3^2/J3^2 = 1.0$ |
| | | |
| d(Mo-Mo) (Å) | 5.695 | 5.6908 |
| $J4^{(c)}$ (Å$^{-1}$) | $J4^1$= -0.0002 AFM | $J4^1$= -0.0002 AFM |
| $j(X)^d$ (Å$^{-1}$) | $j$(Te2): -0.0027x2 | $j$(Te2): -0.0027x2 |
| ($\Delta h(X)^e$Å, $l_n'/l_n^f$, MoXMo$^g$) | (-0.190, 2.19, 104.20°) | (-0.190, 2.19, 104.20°) |
| $j(X)^d$ (Å$^{-1}$) | $j$(Te1): 0.0026x2 | $j$(Te1): 0.0026x2 |
| ($\Delta h(X)^e$Å, $l_n'/l_n^f$, MoXMo$^g$) | (0.393, 4.62, 82.20°) | (0.393, 4.62, 82.20°) |
| $J$n/$J$max | $J4^1/J3^2 = 0.002$ | $J4^2/J3^2 = 0.002$ |
| | | |
| d(Mo-Mo) (Å) | 6.330 | 6.330 |
| $J5^{(c)}$ (Å$^{-1}$) | $J5^1$= -0.0462 AFM | $J5^2$= -0.0460 AFM |
| $j(X)^d$ (Å$^{-1}$) | $j$(Te1): 0.0476 | $j$(Te4): 0.0472 |
| ($\Delta h(X)^e$Å, $l_n'/l_n^f$, MoXMo$^g$) | (-0.867, 1.78, 132.32°) | (-0.856, 1.57, 131.9°) |
| $J$n/$J$max | $J5^1/J3^2 = 0.48$ | $J5^1/J3^2 = 0.47$ |
| | | |
| d(Mo-Mo) (Å) | 6.581 | 6.575 |
| $J6^{(c)}$ (Å$^{-1}$) | $J6^1$ = -0.0760 AFM | $J6^1$ = -0.0741 AFM |
| $j(X)^d$ (Å$^{-1}$) | $j$(Te1): 0.0393x2 | $j$(Te4): 0.0395x2 |
| ($\Delta h(X)^e$ Å, $l_n'/l_n^f$, MoXMo$^g$) | (-0.726, 1.78, 128.52°) | (-0.729, 1.78, 128.6°) |
| $J$n/$J$max | $J6^1/J3^2$ = 0.78 | $J6^2/J3^2$ = 0.76 |
| | | |
| d(Mo-Mo) (Å) | 6.938 | 6.938 |
| $J7^1(J2^1{}_2)$ (Å$^{-1}$) | $J7^1(J2^1{}_2)$ = -0.0088 AFM | $J7^2(J2^2{}_2)$ = -0.0088 AFM |
| $j(X)^d$ (Å$^{-1}$) | $j$(Mo1): -0.0270 | $j$(Mo2): -0.0270 |
| ($\Delta h(X)^e$ Å, $l_n'/l_n^f$, MoXMo$^g$) | (-0.650, 1.0, 180°) | (-0.650, 1.0, 180°) |
| $j(X)^d$ (Å$^{-1}$) | $j$(Te2): 0.0204 | $j$(Te3): 0.0208 |
| ($\Delta h(X)^e$ Å, $l_n'/l_n^f$, MoXMo$^g$) | (0.492, 1.0, 104.18°) | (0.500, 1.0, 104.01°) |
| $J$n/$J$max | $J7^1/J3^2$ = 0.09 | $J7^2/J3^2$ = 0.09 |
| | | |
| d(Mo-Mo) (Å) | 7.218 | 7.218 |
| $J9^1$ (Å$^{-1}$) | $J9^1$ = -0.0113 AFM | $J9^2$ = -0.0117 AFM |
| $J$n/$J$max | $J9^1/J3^2$ = 0.12 | $J9^2/J3^2$ = 0.12 |
| Between layers | | |
| d(Mo1-Mo2) (Å) | $J$n (Å$^{-1}$) | $J$n/$J$max |
| 7.128 | $J8$ = -0.0117 AFM | 0.12 |
| 7.241 | $J10$ = -0.0106 AFM | 0.11 |
| 7.261 | $J11$ = -0.0079 AFM | 0.08 |
| 7.598 | $J12$ = -0.0622 AFM | 0.64 |
| 7.640 | $J13$ = -0.0102 FM | 0.11 |

| | 7.987 | $J14$ = -0.0102 AFM ↔ -0.0703 AFM (l/l=1.98) | 0.11↔0.72 |

[a]XDP: X-ray diffraction from a powder crystal, XDS: X-ray diffraction from a single crystal.
[b]The refinement converged to the residual factor ($R$) values.
[c]$Jn$ in Å$^{-1}$: the magnetic couplings ($Jn < 0$, AFM; $Jn > 0$, FM)
[d]$j(X)$: contributions of the intermediate ion X to the AFM ($j(X) <0$) and FM ($j(X)>0$) components of the coupling $Jn$.
[e]$\Delta h(X)$: the degree of overlapping of the local space between magnetic ions by the intermediate ion X.
[f]$l_n'/l_n$: the asymmetry of the position of the intermediate ion X relative to the middle of the Mo$_i$–Mo$_j$ bond line.
[g]M$_i$XM$_j$: bonding angle.
[h]Small $j(X)$ contributions are not shown.

Table 4. **Crystallographic characteristics and parameters of magnetic couplings ($Jn$) calculated on the basis of structural data and respective distances between magnetic Mn$^{2+}$ (Fe$^{2+}$) ions in the shigaite NaAl$_3$Mn$_6$(SO$_4$)$_2$(OH)$_{18}$12(H$_2$O) and NaAl$_3$Fe$_6$(SO$_4$)$_2$(OH)$_{18}$12(H$_2$O)**

| Crystallographic and magnetic parameters | NaAl$_3$Mn$_6$(SO$_4$)$_2$(OH)$_{18}$ · 12(H$_2$O) [33] Min Name: Shigaite (Data for ICSD-82492) Space group $R$ -3 (N148) $a = b = 9.512$, $c = 33.074$ Å $\alpha = \beta = 90º$, $\gamma = 120º$ Z = 3 Method[a]: XDS (293 K); $R$-value[b] = 0.042 | NaAl$_3$Mn$_6$(SO$_4$)$_2$(OH)$_{18}$ · 12(H$_2$O) Compression of parameters by 0.2 Space group $R$ -3 (N148) $a = b = 9.312$, $c = 32.874$ Å $\alpha = \beta = 90º$, $\gamma = 120º$ Z = 3 | NaAl$_3$Fe$_6$(SO$_4$)$_2$(OH)$_{18}$ ·12(H$_2$O) [34] Min Name: Nikischerite (Data for ICSD-97312) Space group $R$ -3 (N148) $a = b = 9.347$, $c = 33.000$ Å $\alpha = \beta = 90º$, $\gamma = 120º$ Z = 3 Method[a]: XDS (293 K); $R$-value[b] = 0.064 |
|---|---|---|---|
| d(M-X) (Å) | Mn1: octahedron Mn1 - O3 = 2.172 - O3 = 2.177 - O5 = 2.189 - O5 = 2.199 | Mn1: octahedron Mn1 - O3 = 2.132 - O3 = 2.139 - O5 = 2.151 - O5 = 2.159 | Fe1: octahedron Fe1 - O3 = 2.099 - O3 = 2.129 - O5 = 2.144 - O5 = 2.164 |
| | - O4 = 2.205 | - O4 = 2.166 | - O4 = 2.146 |
| | - O4 = 2.209 | - O4 = 2.177 | - O4 = 2.156 |
| d(M-M) (Å) | 3.163 | 3.096 | 3.116 |
| $J1^{(c)}$ (Å$^{-1}$) | $J1$ = 0.0558 (FM) | $J1$ = 0.0492 (FM) | $J1$ = 0.0340 (FM) |
| $j(X)^d$ (Å$^{-1}$) | $j$(O4): 0.0279x2 | $j$(O4): 0.0246x2 | $j$(O4): 0.0170x2 |
| ($\Delta h$(X)$^e$ Å, $l_n$'/$l_n^f$, MXM$^g$) | (0.139, 1.0, 91.55°) | (0.118, 1.0, 91.13°) | (0.083, 1.0, 92.9°) |
| $J$n/$J$max | $J1/J1$ = 1 | $J1/J1$ = 1 | $J1/J1$ = 1 |
| | | | |
| d(M-M) (Å) | 3.175 | 3.108 | 3.115 |
| $J2^{(c)}$ (Å$^{-1}$) | $J2$ = 0.0398 (FM) | $J2$ = 0.0325 (FM) | $J2$ = 0.0243 (FM) |
| $j(X)^d$ (Å$^{-1}$) | $j$(O3): 0.0171 | $j$(O3): 0.0133 | $j$(O5): 0.0182 |
| ($\Delta h$(X)$^e$ Å, $l_n$'/$l_n^f$, MXM$^g$) | (0.086, 1.01, 93.77°) | (0.086, 1.01, 93.40°) | (0.088, 1.02, 92.6°) |
| $j(X)^d$ (Å$^{-1}$) | $j$(O5): 0.0227 | $j$(O5): 0.0227 | $j$(O3): 0.0061 |
| ($\Delta h$(X)$^e$ Å, $l_n$'/$l_n^f$, MXM$^g$) | (0.114, 1.01, 92.69°) | (0.114, 1.01, 92.30°) | (0.136, 1.03, 94.9°) |
| $J$n/$J$max | $J2/J1$ = 0.71 | $J2/J1$ = 0.66 | $J2/J1$ = 0.71 |
| | | | |
| d(M-M) (Å) | 5.471 | 5.356 | 5.379 |
| $J3^{(c)}$ (Å$^{-1}$) | $J3$ = -0.0506 (AFM) | $J3$ = -0.0032 (AFM) | $J3$ = -0.0583 AFM |
| $j(X)^d$ (Å$^{-1}$) | $j$(O3): -0.0328 | $j$(O3): -0.0348 | $j$(O3): -0.0362 |
| ($\Delta h$(X)$^e$ Å, $l_n$'/$l_n^f$, MXM$^g$) | (-0.415, 1.82, 137.53°) | (-0.422, 1.82, 136.97°) | (-0.434, 1.89, 137.23°) |
| $j(X)^d$ (Å$^{-1}$) | $j$(O4): -0.0261 | $j$(O4): -0.0278 | $j$(O4): -0.0296 |
| ($\Delta h$(X)$^e$ Å, $l_n$'/$l_n^f$, MXM$^g$) | (-0.328, 1.84, 134.08°) | (-0.335, 1.84, 133.48°) | (-0.356, 1.84, 134.08°) |
| $j(X)^d$ (Å$^{-1}$) | $j$(O4): 0.0043 | $j$(O4): 0.0042 | $j$(O4): 0.0040 |

| (Δ*h*(X)[e] Å, *l*$_n$'/*l*$_n$[f], MXM[g]) | (0.507, 3.95, 96.51°) | (0.476, 3.95, 96.29°) | (0.473, 4.14, 95.80°) |
|---|---|---|---|
| *j*(X)[d] (Å$^{-1}$) | *j*(O3): 0.0040 | *j*(O3): 0.0039 | *j*(O3): 0.0035 |
| (Δ*h*(X)[e] Å, *l*$_n$'/*l*$_n$[f], MXM[g]) | (0.467, 3.88, 97.75°) | (0.436, 3.88, 97.55°) | (0.412, 4.07, 97.56°) |
| *j*(X)[d] (Å$^{-1}$) |  | *j*(Mn): 0.0513 (FM) |  |
| (Δ*h*(X)[e] Å, *l*$_n$'/*l*$_n$[f], MXM[g]) |  | (0.735, 1.01, 119.39°) |  |
| *J*n/*J*max | *J*3/*J*1 = -0.91 | *J*3/*J*1 = -0.07 | *J*3/*J*1 = -1.71 |
|  |  |  |  |
| d(M-M) (Å) | 5.499 | 5.383 | 5.396 |
| *J*4[(c)] (Å$^{-1}$) | *J*4 = -0.0503 (AFM)[h] | *J*4 = -0.0043 (AFM)[h] | *J*4 = -0.0521FM)[h] |
| *j*(X)[d] (Å$^{-1}$) | *j*(O5): -0.0293 | *j*(O5): -0.0311 | *j*(O5): -0.0297 |
| (Δ*h*(X)[e] Å, *l*$_n$'/*l*$_n$[f], MXM[g]) | (-0.371, 1.84, 135.9°) | (-0.377, 1.84, 135.3°) | (-0.359. 1.88, 134.5) |
| *j*(X)[d] (Å$^{-1}$) | *j*(O5): -0.0294 | *j*(O5): -0.0313 | *j*(O5): -0.0304 |
| (Δ*h*(X)[e] Å, *l*$_n$'/*l*$_n$[f], MXM[g]) | (-0.375, 1.83, 136.1°) | (-0.313, 1.83, 135.5°) | (-0.304, 1.83, 135.3°) |
| *j*(X)[d] (Å$^{-1}$) | *j*(O5): 0.0041 | *j*(O5): 0.0040 | *j*(O5): 0.0041 |
| (Δ*h*(X)[e] Å, *l*$_n$'/*l*$_n$[f], MXM[g]) | (0.477, 3.87, 97.7°) | (0.446, 3.88, 97.6°) | (0.474, 4.0, 96.8°) |
| *j*(X)[d] (Å$^{-1}$) | *j*(O5): 0.0042 | *j*(O5): 0.0041 | *j*(O5): 0.0039 |
| (Δ*h*(X)[e] Å, *l*$_n$'/*l*$_n$[f], MXM[g]) | (0.487, 3.88, 97.5°) | (0.455, 3.88, 97.3°) | (0.452, 4.0, 97.1°) |
| *j*(X)[d] (Å$^{-1}$) |  | *j*(Mn): 0.0500 (FM) |  |
| (Δ*h*(X)[e] Å, *l*$_n$'/*l*$_n$[f], MXM[g]) |  | (0.724, 1.00, 120.0°) |  |
| *J*n/*J*max | *J*4/*J*1 = -0.90 | *J*4/*J*1 = -0.09 | *J*4/*J*1 = -1.53 |
|  |  |  |  |
| d(M-1) (Å) | 5.505 | 5.389 | 5.415 |
| *J*5[(c)] (Å$^{-1}$) | *J*5 = -0.0499 (AFM) | *J*5 = -0.0052 (AFM) | *J*5 = -0.0567 (AFM) |

| $j(X)^d$ (Å$^{-1}$) | $j$(O3): -0.0326 | $j$(O3): -0.0345 | $j$(O3): -0.0362 |
|---|---|---|---|
| ($\Delta h(X)^e$ Å, $l_n$'/$l_n^f$, MXM$^g$) | (-0.414, 1.85, 137.53°) | (-0.420, 1.85, 136.98°) | (-0.445, 1.84, 138.13°) |
| $j(X)^d$ (Å$^{-1}$) | $j$(O4): -0.0255 | $j$(O4): -0.0271 | $j$(O4): -0.0282 |
| ($\Delta h(X)^e$ Å, $l_n$'/$l_n^f$, MXM$^g$) | (-0.322, 1.82, 134.04°) | (-0.329, 1.85 133.45) | (-0.343, 1.88, 134.0°) |
| $j(X)^d$ (Å$^{-1}$) | $j$(O4): 0.0043 | $j$(O4): 0.0042 | $j$(O4): 0.0040 |
| ($\Delta h(X)^e$ Å, $l_n$'/$l_n^f$, MXM$^g$) | (0.503, 3.91, 97.06°) | (0.472, 3.90, 96.84°) | (0.475, 4.09, 96.3°) |
| $j(X)^d$ (Å$^{-1}$) | $j$(O3): 0.0039 | $j$(O3): 0.0037 | $j$(O3): 0.0037 |
| ($\Delta h(X)^e$ Å, $l_n$'/$l_n^f$, MXM$^g$) | (0.449, 3.84, 98.66°) | (0.418, 3.84, 98.47°) | (0.436, 4.02, 97.45°) |
| $j(X)^d$ (Å$^{-1}$) |  | $j$(Mn): 0.0487 (FM) |  |
| ($\Delta h(X)^e$ Å, $l_n$'/$l_n^f$, MXM$^g$) |  | (0.707, 1.00, 120.60°) |  |
| $J$n/$J$max | $J5/J1$ = -0.89 | $J5/J1$ = -0.11 | $J5/J1$ = -1.67 |
|  |  |  |  |
| d(M1-M1) (Å) | 6.337 | 6.204 | 6.232 |
| $J6^{(c)}$ (Å$^{-1}$) | $J6^*$ = 0.0485 (FM) | $J6^*$ = 0.0454 (FM) | $J6^*$ = 0.0488 (FM) |
| $j(X)^d$ (Å$^{-1}$) | $j$(O4): 0.0273 | $j$(O4): 0.0268 | $j$(O4): 0.0281 |
| ($\Delta h(X)^e$ Å, $l_n$'/$l_n^f$, MnXM$^g$) | (0.548 1.01, 116.64°) | (0.515 1.01, 116.61°) | (0.545, 1.01, 117.62°) |
| $j(X)^d$ (Å$^{-1}$) | $j$(O3): 0.0239 | $j$(O3): 0.0233 | $j$(O3): 0.0239 |
| ($\Delta h(X)^e$ Å, $l_n$'/$l_n^f$, MXM$^g$) | (0.480, 1.02, 118.54°) | (0.448, 1.02, 118.44°) | (0.486, 1.0, 116.04°) |
| $J$n/$J$max | $J6/J1$ = 0.87 | $J6/J1$ = 0.92 | $J6/J1$ = 1.44 |
|  |  |  |  |
| d(M-M) (Å) | 6.349 | 6.216 | 6.231 |
| $J7^{(c)}$ (Å$^{-1}$) | $J7^*$ = 0.0468 (FM) | $J7^*$ = 0.0448 (FM) | $J7^*$ = 0.0490 (FM) |
| $j(X)^d$ (Å$^{-1}$) | $j$(O5): 0.0254x2 | $j$(O5): 0.0248x2 | $j$(O5): 0.0264x2 |

| ($\Delta h(X)$[e] Å, $l_n'/l_n$[f], MXM[g]) | (0.511 1.0, 117.91°) | (0.479. 1.0, 117.69°) | (0.512 1.0, 126.53°) |
|---|---|---|---|
| $J$n/$J$max | $J7/J1 = 0.84$ | $J7/J1 = 0.91$ | $J7/J1 = 1.44$ |
| *Interplane couplings* | | | |
| d(M-M) (Å) | 11.025 | 10.958 | 11.00 |
| $J$inter plane (Å$^{-1}$) | $J$inter = 0.0045 (FM) | $J$inter = 0.0025 (FM) | $J$inter = 0.0015 (FM) |

[a]XDS: X-ray diffraction from a single crystal.
[b]The refinement converged to the residual factor ($R$) values.
[c]$J$n in Å$^{-1}$: the magnetic couplings ($J$n < 0, AFM; $J$n > 0, FM)
[d]$j(X)$: contributions of the intermediate ion X to the AFM ($j(X) < 0$) and FM ($j(X) > 0$) components of the coupling $J$n.
[e]$\Delta h(X)$: the degree of overlapping of the local space between magnetic ions by the intermediate ion X.
[f]$l_n'/l_n$: the asymmetry of the position of the intermediate ion X relative to the middle of the $M_i$–$M_j$ bond line.
[g]$M_i X M_j$: bonding angle.
[h]Small $j(X)$ contributions are not shown.